\documentclass[aps,prx,reprint,superscriptaddress,longbibliography,floatfix]{revtex4-2}
\usepackage{lmodern,microtype,capt-of}
\usepackage[T1]{fontenc}
\usepackage{amsmath,amssymb,amsthm,bm,booktabs,graphicx}
\usepackage[colorlinks=true,linkcolor=blue,citecolor=blue,urlcolor=blue]{hyperref}
\newcommand{\Tr}{\operatorname{Tr}}
\newcommand{\ket}[1]{|#1\rangle}
\newcommand{\bra}[1]{\langle#1|}
\newcommand{\E}{\mathcal E}
\newcommand{\id}{\mathcal I}
\newcommand{\pos}[1]{[#1]_+}
\theoremstyle{definition}
\newtheorem{theorem}{Theorem}
\newtheorem{corollary}{Corollary}
\newtheorem{proposition}{Proposition}

\begin{document}
\title{Quantum Codes for Distributed Quantum Fault Sensing}
\author{Xiu-Hao Deng}
\email{dengxiuhao@iqasz.cn}
\affiliation{International Quantum Academy, Shenzhen, 518048, China}
\affiliation{Shenzhen Branch, Hefei National Laboratory, Shenzhen, 518048, China}

\begin{abstract}
Targeted recovery in modular quantum processors requires identifying the component affected by a fault. Independent qubits can monitor faults through repeated preparation and measurement. Bell-pair probes improve the guaranteed correct-site probability for unknown local channels, but require partners kept away from the fault. Can entanglement among exposed probes supply this advantage without protected storage? We establish a quantum-code criterion for optimal distributed fault sensing of one unknown local qubit channel with strength in a prescribed positive range. For one interrogation with zero false alarms and no background noise, suitable code states attain the optimal worst-case correct-site probability, improving on independent probes by 50\% with every qubit exposed. Exact attainment is equivalent to mutually orthogonal input and single-qubit Pauli error subspaces. Pairwise entanglement gives no improvement, and entanglement depth at most four cannot exceed 95.54\% of the optimum without protected ancillas. Approaching the optimum without protected ancillas also requires a number of probes proportional to the array size to participate in larger entangled factors. For a persistent fault under memoryless depolarizing background of fixed strength $0<\eta<1$, independent probes already attain the optimal logarithmic interrogation scaling at fixed success between zero and one and a positive false-alarm budget. In three 1000-site comparisons with specified readouts, an encoded-probe protocol certifies fewer interrogations than independent probes and only one more than optimally decoded ideal Bell pairs, while storing no protected qubits.
\end{abstract}
\maketitle
\raggedbottom

\begin{figure*}[t]
\centering
\includegraphics[width=\textwidth]{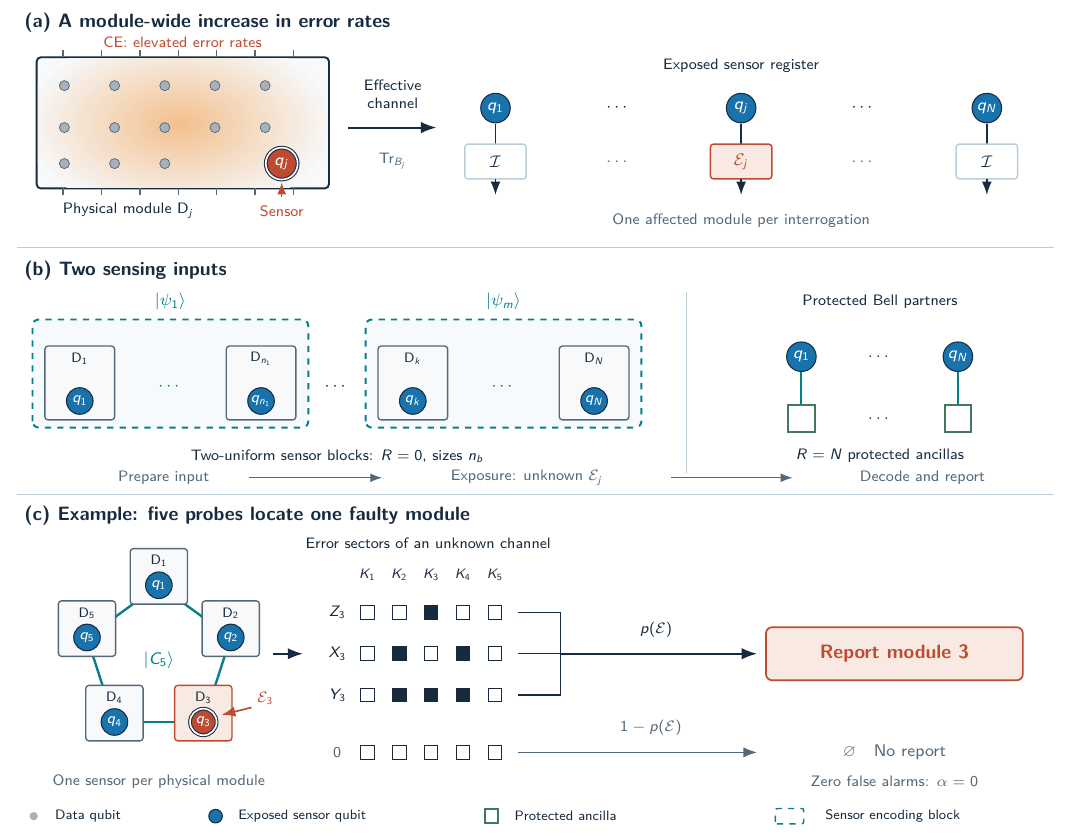}
\caption{\textbf{An illustrative modular realization of encoded fault sensing.}
(a) Orange shading denotes a module-wide increase in error rates, rather than a particular pattern of simultaneous errors. A catastrophic error (CE) describes its impact on error correction, without specifying a cause or requiring conditionally correlated errors. Under the conditions in Appendix~\ref{app:physical}, an affected module induces an unknown sensor channel $\E_j$; the other exposed sensor channels are identity in the ideal model.
(b) Two sensing inputs: blocks of exposed sensors across physical modules whose one- and two-sensor reduced states are maximally mixed (two-uniform states), or exposed--protected Bell pairs. Dashed boundaries enclose sensor blocks; solid boundaries enclose physical modules. Data qubits are outside the sensor encoding. The time strip places the unknown channel between preparation and decoding. Bell partners bypass the fault and require $R=N$ retained qubits.
(c) A five-cycle illustrates the readout of a state satisfying the general quantum-code criterion. The three Pauli error sectors at site 3 are readout sectors of an arbitrary unknown single-qubit channel, and all report module 3. Their total probability is $p(\E)$. The zero syndrome gives no report with probability $1-p(\E)$; silence does not certify absence of a fault. Filled and empty squares denote stabilizer eigenvalues $-1$ and $+1$. Graph edges specify the input state without prescribing hardware connectivity.}
\label{fig:mechanism}
\end{figure*}

\section{Introduction}

Targeted recovery in modular quantum processors requires knowing which component has been affected by a fault. An outer code can treat an identified faulty module as an erasure~\cite{Xu2022}, while spatial information about error bursts can guide code deformation and decoder updates~\cite{Suzuki2022}. Location information can therefore support recovery before the microscopic fault mechanism has been fully characterized.

Superconducting processors illustrate the problem. Radiation-induced quasiparticles affect coherence~\cite{Vepsalainen2020}, and spatially extended relaxation and error bursts have been observed~\cite{Wilen2021,McEwen2022,Harrington2025,Li2025}. Rare bursts also limit high-distance repetition codes in processors with below-threshold surface-code memories; some burst mechanisms remain unidentified~\cite{Google2025}. We use \emph{catastrophic error} (CE)~\cite{McEwen2022,Xu2022} for an excursion that overwhelms a device's error-correction scheme, regardless of its microscopic cause. Error rates can rise across a module even when physical errors remain conditionally independent; whether increasing code distance helps depends on the burst and decoder~\cite{Tan2024}.

Repeated local measurements already reveal such disturbances. McEwen et al.~\cite{McEwen2022} used repeated excited-state preparation, idling and local measurement to resolve relaxation bursts in space and time. Varying preparation and measurement bases extends sensitivity beyond relaxation. Code syndromes~\cite{Xu2022,Suzuki2022} and environmental monitors~\cite{Orrell2021} also supply fault information. Repeated error detection supports in situ calibration~\cite{Kelly2016}, while syndrome statistics yield noise estimates and, under identifiability conditions, Pauli-channel reconstruction~\cite{Wagner2021,Wagner2022}. Scalable characterization reveals processor-wide noise correlations~\cite{Harper2020}. We ask how much entanglement can improve fault identification with a fixed array of exposed probes.

Quantum sensing uses coherence and correlations to extract information~\cite{Degen2017}; in a network, their benefit depends on the task~\cite{Proctor2018,Eldredge2018}. We consider an array on which one unknown local qubit channel acts, with strength in a prescribed positive range. The detector identifies the affected site without reconstructing the channel, while limiting false alarms when no fault is present. In a modular realization, each module contributes one sensor to a register separate from the data. Under the conditions in Appendix~\ref{app:physical}, a disturbance confined to one module induces one unknown sensor channel.

This task fits within channel-position finding~\cite{Zhuang2020}. For known qubit faults, the best strategy depends on the channel~\cite{Skotiniotis2024}, and idler-free protocols exploit multipartite probe correlations~\cite{Pereira2021,Harney2021,HarneyDynamic2021}. Retained ancillas offer another resource for discrimination~\cite{Piani2009}, as illustrated by the idler that bypasses the target in quantum illumination~\cite{Tan2008}. Here a \emph{protected ancilla} bypasses the unknown channel and remains available throughout interrogation, with noiseless storage in the ideal model. We compare independent probes, exposed--protected Bell pairs and entangled exposed probes under the same unknown-channel and false-alarm assumptions. The independent baseline tests the gain from input entanglement; the Bell baseline tests whether optimal performance requires protected storage.

A quantum code offers a way to make local faults distinguishable. The Knill--Laflamme conditions characterize correctable noise~\cite{Knill1997}; nondegenerate codes separate local-error sectors~\cite{Laflamme1996}, with an established connection to maximally mixed low-order marginals~\cite{Scott2004}. For the code to act as a sensor, those sectors must carry location information. Correctability alone need not distinguish sites, and the sensing task does not require recovery of arbitrary logical information. The resource cost of this structure connects our question to metrological entanglement-depth bounds~\cite{Hyllus2012,Toth2012} and to the roles of coding and ancillary systems in metrology~\cite{Kessler2014,Zhou2018,Layden2019} and communication~\cite{Brun2006}. Figure~\ref{fig:mechanism} shows the modular setting and the encoded and protected-ancilla architectures.

We establish a necessary-and-sufficient quantum-code criterion for attaining the optimal correct-site probability in a single interrogation with zero false alarms and no background noise: the input support and all single-site Pauli error supports must be mutually orthogonal. Families of suitable code states attain this optimum with every probe exposed, improving on independent probes by 50\% and matching ideal protected Bell pairs. We reduce bounded-depth optimization to cluster discrimination and resource allocation, obtaining the exact depth-two frontier and explicit gains from larger blocks. Finite-error bounds quantify the cost of missing protected ancillas and show that approaching the optimum without them requires a number of probes proportional to the array size to participate in larger entangled factors. These bounds allow unrestricted global readout.

Background noise shifts the comparison to repeated interrogation. Channel-divergence bounds constrain strategies with retained ancillas and adaptive control~\cite{Wilde2020}. For a persistent site and channel under memoryless depolarization of fixed strength $0<\eta<1$, logarithmically many interrogations are necessary as the array grows at fixed success probability between zero and one and a positive false-alarm budget. Independent probes attain this order, so the remaining comparison concerns finite round counts and resource costs. For specified readouts, encoded probes can reduce the interrogation count relative to independent probes while approaching ideal Bell performance without protected storage. We include preparation and measurement noise, and examine how additional local sensors and prior channel knowledge alter the comparison.

\section{Results}
\subsection{Sensing task and optimal correct-site probability}

Consider $N\geq1$ exposed qubits and $R\geq0$ protected ancilla qubits, interrogated once in parallel (Fig.~\ref{fig:mechanism}a,b). The no-fault, or null, hypothesis assigns the identity channel to every qubit. Under a \emph{fault}, a nonidentity, completely positive and trace-preserving (CPTP) map $\E$ acts at one unknown exposed site $j$, while all other qubits undergo identity evolution. We quantify the fault strength by its process infidelity relative to the identity channel, $p(\E)=1-F_e(I/2,\E)$. Here $F_e(I/2,\E)$ is the entanglement fidelity for a maximally mixed input: the overlap with the initial Bell state after $\E$ acts on one member of the pair~\cite{Nielsen2002}. In terms of Kraus operators $K_\ell$,
\begin{equation}
 p(\E)=1-\frac14\sum_\ell|\Tr K_\ell|^2.
 \label{eq:strength}
\end{equation}
Thus $p(\E)$ is also the entanglement infidelity for this input. It characterizes the channel, whether or not the sensing protocol uses entangled probes, and is distinct from the probability that a physical event occurs.

In the modular realization of Fig.~\ref{fig:mechanism}a, each physical module contributes one exposed sensor. Only the sensor register is encoded; a disturbance may also affect many data qubits in the same module. If the other module degrees of freedom start in a fixed state uncorrelated with the sensor input, tracing them out yields a local CPTP sensor channel. Appendix~\ref{app:physical} states the reduction and its remaining conditions. A burst spanning several modules instead produces a multi-site fault, outside the ideal task.

Every protocol knows the same allowed channel family: all qubit CPTP maps with $0<p_{\min}\leq p(\E)\leq p_{\max}\leq1$. It knows neither the actual site nor the channel's mechanism and parameters. The site and channel are independent of the protocol's private preparation randomness.

A protocol prepares $\rho_c$ with probability $\lambda_c$ and retains the classical label $c$. Readout uses a conditional positive-operator-valued measure (POVM), $\{M_{0|c},\ldots,M_{N|c}\}$, whose outcomes report a site or give no report (outcome zero). We maximize the worst-case probability $v$ of reporting the correct site, subject to a constraint on the false-alarm probability $F$:
\begin{align}
 F&=\sum_c\lambda_c\sum_{j=1}^N\Tr(M_{j|c}\rho_c)\leq\alpha,\nonumber\\
 v&=\inf_{j,\E}\sum_c\lambda_c\Tr[M_{j|c}\E_j(\rho_c)] .
 \label{eq:task}
\end{align}
Here $0\leq\alpha\leq1$, and $\E_j$ acts as identity on every qubit other than $j$. False alarms are averaged over the preparation labels. The score is unconditional: an incorrect site and no report both count as failure. The endpoint $\alpha=0$ requires silence under the null hypothesis; $\alpha=1$ permits a site report on every outcome.

We count two input resources: the number $R$ of protected ancillas and the entanglement depth. Entanglement depth at most $k$, also called $k$-producibility, means that each conditional state $\rho_c$ can be decomposed into pure states that factor into clusters of at most $k$ qubits. Different components may use different partitions. Cluster size includes both exposed and protected input qubits. All quantum systems retained through exposure enter this accounting; temporary ancillas introduced only for readout are excluded from these input resources.

Readout is unrestricted: every protocol, including the independent-probe baseline, may use an arbitrary global POVM. Entanglement depth constrains only the input correlations; it specifies neither circuit depth nor elapsed time. We write $V_{k,\alpha}(N,R)$ for the largest achievable score. \emph{Distributed} refers to the distinct exposed sites; the model imposes no spatial separation, communication latency or geometric gate constraint.

The Bell and graph-block constructions require quantum readout only within each block, followed by classical aggregation. The converses allow global readout and remain valid under additional measurement restrictions.

Optimizing over all input states and measurements, with unrestricted access to protected ancillas, gives the worst-case, or \emph{minimax}, correct-site probability
\begin{equation}
 B_\alpha=p_{\min}+(1-p_{\min})\alpha/N .
 \label{eq:benchmark}
\end{equation}
The value $B_\alpha$ bounds every protocol, including those with no protected ancillas. Bell pairs provide one way to attain it: pair each exposed probe with a protected ancilla (Fig.~\ref{fig:mechanism}b, right). A channel of strength $p=p(\E)$ produces a Bell-measurement outcome orthogonal to the prepared Bell state with probability $p$, identifying the affected site. If every pair returns its prepared Bell state, report a uniformly random site with probability $\alpha$. This adds $(1-p)\alpha/N$ to the success probability. The minimum over the allowed channel family occurs at $p_{\min}$.

To see why no protocol can do better, choose the allowed isotropic Pauli channel of strength $p_{\min}$. Its identity branch carries no information about location. The false-alarm constraint limits its contribution to site-averaged success to $(1-p_{\min})\alpha/N$; all other branches together contribute at most $p_{\min}$. Worst-site success cannot exceed the site average, giving the upper bound $B_\alpha$. This argument uses the full allowed channel family. Prior knowledge that restricts the family to a specified channel can change the optimum.

The independent-probe comparison measures what is gained by introducing input entanglement under the same channel and false-alarm assumptions. Its optimized value is
\begin{equation}
 V_{1,\alpha}=\frac{2p_{\min}}3+
       \left(1-\frac{2p_{\min}}3\right)\frac{\alpha}{N}.
 \label{eq:independent}
\end{equation}
This value is attained by choosing each probe uniformly from the six Pauli eigenstates and measuring in its preparation basis. Thus the independent strategy already allows different local states and retains their classical preparation labels. It needs no two-qubit gates. The matching converse allows mixed inputs, any number of uncorrelated protected qubits, and global readout (Appendix~\ref{app:responses}). At zero false alarms, the optimal independent score is $2p_{\min}/3$, whereas $B_0=p_{\min}$ is attained by protected Bell pairs. We next determine when correlations among exposed probes close this performance gap without the Bell partners' protected storage. These are two resource comparisons within the same sensing task.

\subsection{Quantum-code criterion for optimal fault sensing}

To attain the benchmark with zero false alarms, a sensor must distinguish the undisturbed input support from the error supports associated with different sites. The support of a density operator is the subspace spanned by its eigenvectors with nonzero eigenvalues; an error support is its image under the corresponding Pauli operator. We can express the requirement without choosing a code or a block size. Set $P_0=I$ and $P_{j,a}=\sigma_a^{(j)}\otimes I_A$, where $a=X,Y,Z$ and $A$ denotes the protected ancillas. We work here in the single-interrogation model with identity background.

\begin{theorem}[Support criterion at zero false alarms]\label{thm:support}
Fix an input state $\rho$ and let $Q$ project onto its support. There exists a readout attaining $v=B_0=p_{\min}$ with $F=0$ if and only if
\begin{equation}
 QP_\mu^\dagger P_\nu Q=\delta_{\mu\nu}Q
 \quad(\mu,\nu\in\{0\}\cup\{(j,a)\}).
 \label{eq:support}
\end{equation}
For a retained preparation label, the criterion applies separately to every conditional state of positive probability, with a conditional readout. In particular, a pure input of exposed probes alone attains the benchmark if and only if all its one- and two-probe reduced states are maximally mixed.
\end{theorem}

Attaining the isotropic-channel benchmark requires every single-site Pauli error to be assigned to its site with probability one, while the undisturbed support produces no report. Perfect site identification separates the error supports of different sites. Orthogonality to the undisturbed support, together with the Pauli multiplication relations, also separates different axes at the same site. These requirements give the full support criterion; Appendix~\ref{app:support_proof} gives the proof.

Conversely, the orthogonal projectors $Q_\mu=P_\mu QP_\mu^\dagger$ directly specify a site readout:
\begin{equation}
 M_j=\sum_{a=X,Y,Z}Q_{j,a},\qquad
 M_0=I-\sum_jM_j .
 \label{eq:general_readout}
\end{equation}
To evaluate its response to an arbitrary local channel, write each Kraus operator as $K_\ell=a_\ell I+\sum_a b_{\ell a}\sigma_a$. Then
\begin{equation}
 \Pr(j\mid\E_j)=\sum_{\ell,a}|b_{\ell a}|^2=p(\E),
 \qquad \Pr(0\mid\E_j)=1-p(\E).
 \label{eq:general_response}
\end{equation}
Every other site has zero reporting probability. Orthogonality resolves the error contributions even for coherent and nonunital channels, so this calculation places no Pauli-mixture assumption on $\E$.

\begin{figure*}[t]
\centering
\includegraphics[width=\textwidth]{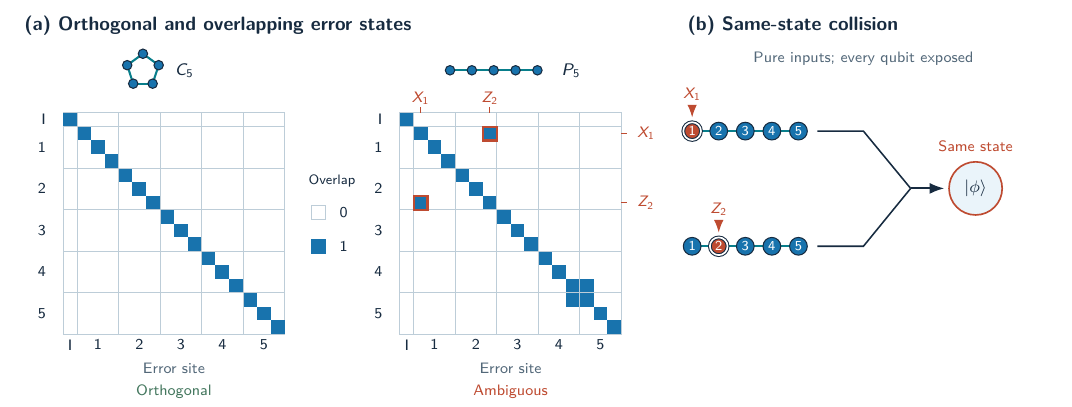}
\caption{\textbf{Orthogonal error states distinguish fault locations.}
(a) Exact squared overlaps $G_{\mu\nu}=|\langle\psi|P_\mu^\dagger P_\nu|\psi\rangle|^2$ for the five-cycle $C_5$ and open path $P_5$. The five-cycle separates all sixteen states. These are pure inputs with every qubit exposed; general mixed inputs require the support criterion in Eq.~\eqref{eq:support}. Red labels and outlined cells mark the path's identical states produced by $X_1$ and $Z_2$. Rows and columns follow $I,X_1,Y_1,Z_1,\ldots,X_5,Y_5,Z_5$, grouped by site; white denotes zero overlap and blue unit overlap.
(b) For the fixed path input, $|\phi\rangle=X_1|P_5\rangle=Z_2|P_5\rangle$: faults at different sites produce the same state, so no measurement can distinguish their locations. Edges specify graph states, not a required hardware coupling geometry. The squared-overlap plots visualize exact orthogonality; the calibrated approximate construction uses the complex Gram matrix, including its phases.}
\label{fig:distinguishability}
\end{figure*}

Equation~\eqref{eq:support} is the orthogonal, nondegenerate case of the Knill--Laflamme error-correction condition~\cite{Knill1997}. Degeneracy is compatible with error correction, but can erase the distinction between fault locations. Optimal sensing in the present task therefore requires this stronger form, even though the sensor need not recover arbitrary logical information. Counting the dimensions of the orthogonal error supports gives the nondegenerate quantum Hamming packing bound, restricted to errors on exposed qubits. If $K=\operatorname{rank}\rho$, then
\begin{equation}
 K(1+3N)\leq 2^{N+R}.
 \label{eq:packing}
\end{equation}
For a pure input with every qubit exposed, Eq.~\eqref{eq:support} reduces to two-uniformity: every one- and two-qubit reduced state is maximally mixed~\cite{Scott2004}. Equivalently,
\begin{equation}
 \langle\Psi|P_\mu^\dagger P_\nu|\Psi\rangle=\delta_{\mu\nu}.
 \label{eq:general_sectors}
\end{equation}
Figure~\ref{fig:distinguishability} contrasts orthogonal error states with faults at different sites that produce the same state.

For a mixed input, the support carries information that the two-probe marginals do not capture. A fully mixed array has maximally mixed marginals, yet is unchanged by a Pauli fault and conveys no location information. A mixture can nevertheless satisfy Eq.~\eqref{eq:support} if distinct Pauli error images of its support, including the identity image, are mutually orthogonal. In coding terminology, the code is then pure with respect to this error set; this does not mean that the density operator is a pure state. Attaining the benchmark requires this support condition; the input itself need not be pure.

When the error states are not exactly orthogonal, a readout calibrated to a known pure input can still give a correct-site guarantee. Appendix~\ref{app:approximate} gives this construction and a separate averaged necessary condition for near-optimal performance.

\paragraph*{Code-state constructions.}

For a pure stabilizer state, two-uniformity holds precisely when every nonidentity element of the full stabilizer group has Pauli weight at least three, meaning that it acts nontrivially on at least three qubits. This lets us test Eq.~\eqref{eq:general_sectors} using the stabilizer group and obtain the readout from stabilizer measurements. Cycle graph states $C_n$ with $n\geq5$ meet the criterion~\cite{Hein2006}. Their generators are $K_i=Z_{i-1}X_iZ_{i+1}$, with cyclic indices.

The construction extends to large arrays using blocks of bounded size. For $5\mid N$, a product of five-cycles attains the benchmark at $R=0$ and depth five. Combining five- and six-cycles covers every $N\geq20$ at depth at most six. Thus, the required correlations can remain confined to small blocks as the array grows.

\paragraph*{A five-qubit example.}

The five-cycle $C_5$ makes the mechanism explicit (Fig.~\ref{fig:mechanism}c); its overlap matrix in Fig.~\ref{fig:distinguishability}a displays the required orthogonality.

Define the syndrome bit $s_i$ to be zero for stabilizer eigenvalue $+1$ and one for eigenvalue $-1$. In the order $(K_1,\ldots,K_5)$, the three Pauli operators at site 3 produce
\begin{equation}
 Z_3:\ 00100,\qquad X_3:\ 01010,\qquad Y_3:\ 01110 .
 \label{eq:syndrome_example}
\end{equation}
The three patterns differ, but each identifies the same affected probe (Fig.~\ref{fig:mechanism}c). Cyclic shifts give the records for the other sites. All fifteen single-site Pauli syndromes are distinct and nonzero. The decoder groups them by location; resolving the Pauli axis supplies more information than identifying the affected site requires.

The three recognized sectors at the affected site have total probability $p(\E)$; the null syndrome has probability $1-p(\E)$ (Fig.~\ref{fig:mechanism}c). At zero false alarms every reported location is correct, but silence remains possible under a fault. Randomizing on a null record as in the Bell protocol gives a response that depends on the channel only through its strength.

For an actual fault of strength $p=p(\E)$, the three unconditional decision probabilities are
\begin{equation}
 \begin{aligned}
 P_{\rm correct}&=p+(1-p)\alpha/N,\\
 P_{\rm wrong}&=(1-p)\alpha(1-1/N),\\
 P_{\rm none}&=(1-p)(1-\alpha).
 \end{aligned}
 \label{eq:decision_probabilities}
\end{equation}
Under the no-fault hypothesis, the total reporting probability is $\alpha$.

For a specified Pauli fault with identity weight $1-p$, the common identity branch bounds the success probability by $p+(1-p)/N$ when a site must always be reported ($\alpha=1$ in our convention). A two-uniform exposed input attains this bound by reporting the resolved error location and guessing uniformly on the null sector. Cycle states therefore attain the assisted bound without ancillas for $N\geq5$, addressing the attainment question posed for rank-three Pauli faults in Ref.~\cite{Skotiniotis2024}. The conclusion concerns Pauli faults; the known-channel optimum for a general non-Pauli fault can differ.

Entanglement alone does not ensure this readout property. For an open-path graph state, the end stabilizer $X_1Z_2$ implies
\begin{equation}
 X_1\ket{G_{\rm path}}=Z_2\ket{G_{\rm path}} .
 \label{eq:path_collision_main}
\end{equation}
The off-diagonal unit overlaps in Fig.~\ref{fig:distinguishability}a expose the ambiguity: two fault locations produce the same full output state, so even a global measurement cannot distinguish them. Mixing each Pauli error with the same identity component preserves the ambiguity for channels in the allowed channel family. This conclusion concerns the fixed path input. Other states on path-connected hardware, and randomized protocols with retained labels, are governed by the support criterion in Eq.~\eqref{eq:support}; their ability to recover logical information alone does not determine whether the affected site can be identified.

\subsection{Finite-error protection cost}

Small blocks can replace protected ancillas only if their error sectors are sufficiently distinguishable. To quantify this restriction, define the benchmark shortfall $\Delta_{k,\alpha}=B_\alpha-V_{k,\alpha}(N,R)$ and the constants
\begin{equation}
 C_4=12+6\sqrt3\simeq22.3923,\qquad
 \kappa_\alpha=\min(1,6\alpha).
 \label{eq:converse_constant}
\end{equation}

\begin{theorem}[Protection cost at bounded depth]\label{thm:protection_main}
For the single-fault task defined above, the following bounds hold for every $N\geq1$, integer $R\geq0$, $k\in\{2,3,4\}$ and $0\leq\alpha\leq1$:
\begin{align}
 \Delta_{k,\alpha}&\geq
   \frac{p_{\min}}{C_4N}\pos{N-R-\kappa_\alpha},\nonumber\\
 \Delta_{k,\alpha}&\leq
   \frac{p_{\min}}{3N}\pos{N-R-\alpha}.
 \label{eq:law}
\end{align}
The subscript $+$ denotes the positive part. The lower bound includes mixed inputs, retained preparation labels and arbitrary global readout. The upper bound is supplied by an explicit depth-two protocol.
\end{theorem}

The upper bound is constructive: randomly assign the available ancillas to exposed sites and use independent probes elsewhere (Appendix~\ref{app:responses}). The resulting input has depth two. Both bounds scale with the number of missing ancillas, up to the stated offsets. Section~\ref{sec:small_blocks} determines the exact depth-two optimum at zero false alarms and compares bounds with explicit constructions at depths three and four.

\begin{figure*}[tbp]
\centering
\includegraphics[width=\textwidth]{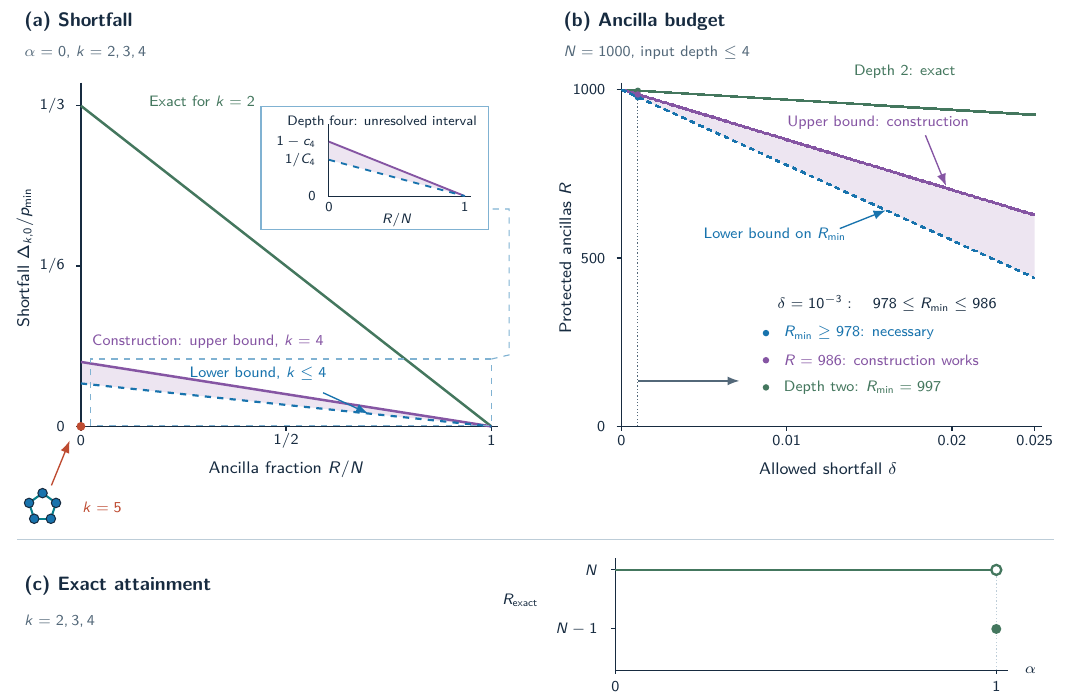}
\caption{\textbf{From performance bounds to a protected-ancilla budget.}
(a) Optimal benchmark shortfall at $\alpha=0$, divided by $p_{\min}$. Green is exact at depth two. The blue lower bound applies at depths two to four. Purple is an upper bound on optimal shortfall from the tetrahedron-block construction at depth four, attained when $4$ divides $N-R$. The shaded band and inset bracket the depth-four optimum between this construction and the converse. The depth-three optimum is not determined. Orange marks the zero-shortfall depth-five construction for $5\mid N$.
(b) Ancilla budgets for $N=1000$, with integer and block-size rounding. Blue is a necessary lower bound on the minimum budget at input depth at most four. Purple is a sufficient upper bound from the best allocation in the proven depth-four construction library, including shared-ancilla blocks. Green is exact at depth two. At $\delta=\epsilon/p_{\min}=10^{-3}$, every depth-at-most-four protocol needs $R\geq978$, while a construction works with $R=986$: $978\leq R_{\min}\leq986$. Depth two needs exactly 997. The inset states these different roles explicitly.
(c) Exact attainment at depths two to four requires $R_{\rm exact}=N$ for $0\leq\alpha<1$, and $N-1$ only at $\alpha=1$. The open upper marker excludes the endpoint; the filled lower marker gives its attainable value. This depth restriction is essential: depth-five blocks can attain the benchmark without protected ancillas. All curves evaluate analytical bounds or constructions.}
\label{fig:resource}
\end{figure*}

At zero false alarms,
\begin{equation}
 \frac{p_{\min}}{C_4}\pos{1-\frac RN}
 \leq\Delta_{k,0}\leq
 \frac{p_{\min}}3\pos{1-\frac RN}.
 \label{eq:zeroFA}
\end{equation}
Both bounds are linear in the missing-ancilla fraction (Fig.~\ref{fig:resource}a). Omitting a fixed fraction therefore leaves a performance gap at depth two to four that does not vanish as the array grows. Five-cycle tilings escape this restriction. Nonzero false alarms change the stated converse only through a finite-size offset. The constant $C_4$ follows from the real representation of Pauli error frames used below.

To translate this result into a resource budget, suppose the target is $v\geq B_0-\epsilon$ and write $\delta=\epsilon/p_{\min}$. A necessary ancilla count and a count sufficient for the explicit construction are
\begin{align}
 R&\geq\max\{0,\lceil N(1-C_4\delta)\rceil\}
       &&\text{necessary},\nonumber\\
 R&\geq\max\{0,\lceil N(1-3\delta)\rceil\}
       &&\text{sufficient}.
 \label{eq:integer_costs_main}
\end{align}
Figure~\ref{fig:resource}b sharpens the sufficient count by allocating all the proven small-block constructions described below. For $N=1000$ and $\delta=10^{-3}$, the depth-four minimum obeys $978\leq R_{\min}\leq986$. The upper endpoint uses 985 Bell pairs, three all-exposed tetrahedron blocks, and one tetrahedron block with three exposed qubits and one protected qubit. Its normalized shortfall is $0.0009539464$. This is the best allocation of the stated construction library. Depth two requires exactly 997 ancillas.

At $\delta=10^{-2}$, the depth-four bracket is $777\leq R_{\min}\leq852$. The quantity $\delta$ measures loss from the benchmark, whose success probability itself need not be one.

For a general false-alarm budget, a benchmark shortfall at most $\epsilon$ requires $R\geq N-\kappa_\alpha-C_4N\epsilon/p_{\min}$. The required ancilla count remains proportional to the array size at fixed $\epsilon/p_{\min}<1/C_4$. In particular, at $R=0$ and $\alpha=0$, depth at most four cannot attain the benchmark, whereas five-cycle tilings can. Five is therefore the smallest sufficient block size for arrays with $5\mid N$.

Analyzing the equality conditions gives a sharper result at exact attainment.
\begin{corollary}[Exact ancilla threshold]\label{cor:exact_main}
For $k\in\{2,3,4\}$, the minimum ancilla count attaining $B_\alpha$ is
\begin{equation}
 R_{\rm exact}(N,\alpha)=
 \begin{cases}
 N,&0\leq\alpha<1,\\
 N-1,&\alpha=1.
 \end{cases}
 \label{eq:exact_main}
\end{equation}
\end{corollary}

Figure~\ref{fig:resource}c resolves the endpoint of the trade-off. The last protected ancilla can be omitted only at $\alpha=1$, when every null record may be assigned to the sole site lacking one. The finite-error bound alone does not determine this endpoint. For example, at $R=N-1$ and $1/6\leq\alpha<1$, the lower bound in Eq.~\eqref{eq:law} vanishes even though the optimum remains strictly below $B_\alpha$.

The equality conditions explain why the endpoint requires every null record to be reported. If an ancilla is missing, perfect identification of the site for every Pauli test state forces a site report on the null input as well. Appendix~\ref{app:attainment} proves this claim and shows that finitely many retained preparation labels suffice to attain the optimum. Thus, at fixed $N,R,k,\alpha$, arbitrarily many retained preparation labels cannot eliminate a strictly positive optimal shortfall.

\paragraph*{Origin of the converse.}

The converse rests on a simple obstruction: overlapping states cannot be distinguished perfectly. Fix a pure input $\ket\psi$ and its readout, and let $e_{ja}=1-\bra\psi a_jM_ja_j\ket\psi$ be the failure probability after Pauli operator $a_j$ acts at site $j$. If errors at two sites produce overlapping states, at least one must have a nonzero probability of failing to identify the correct site. For $a,b\in\{X,Y,Z\}$, the quantitative statement is
\begin{equation}
 |\bra\psi a_j b_l\ket\psi|^2\leq2(e_{ja}+e_{lb}),\qquad j\ne l .
 \label{eq:overlap}
\end{equation}
Split the overlap using $M_j$ and $I-M_j$ and apply Cauchy--Schwarz to obtain this inequality for any global POVM.

A stronger bound retains the joint measurement constraint on all three Pauli error states at each site. For an exposed subset $S$ of size $m\geq2$, let $A_2(S)$ be the sum of squared expectation values of its weight-two Pauli operators. POVM normalization then gives
\begin{equation}
 A_2(S)\leq h_m\sum_{j\in S,a}e_{ja},\qquad
 h_m=\frac{(1+\sqrt{m-1})^2}{2}.
 \label{eq:frame_bound_main}
\end{equation}
The Pauli error states at a given site are orthonormal under the real part of the inner product; cross-site overlaps are already real. Here a frame is the matrix whose three columns are the Pauli error vectors at one site. This observation allows the three-column frames to be treated as real isometries. Applying POVM normalization to these frames gives Eq.~\eqref{eq:frame_bound_main}, as proved in Appendix~\ref{app:converse}.

For a cluster of at most four qubits, an excess of exposed probes over protected ancillas forces either an overlap between errors at different sites or an overlap with the undisturbed input. A pure all-exposed four-qubit cluster illustrates the obstruction. Let $A_w$ denote the sum of squared expectation values of weight-$w$ Pauli operators. Purity and positivity require
\begin{equation}
 A_2\geq A_1+2 .
 \label{eq:four_obstruction}
\end{equation}
Its two-qubit reductions therefore cannot all be maximally mixed. The fully orthogonal error states of the five-cycle are unavailable to this smaller cluster. Appendix~\ref{app:converse} derives the inequality and the corresponding constraints for small clusters containing protected ancillas.

For input depth at most four, the cluster constraints convert the ancilla deficit into a lower bound on the summed Pauli-test failure probabilities. Combining them with Eq.~\eqref{eq:frame_bound_main} and writing $E_{\rm err}=\sum_{j,a}e_{ja}$ gives
\begin{equation}
 K_4E_{\rm err}\geq N-R-1,\qquad
 K_4E_{\rm err}+6F\geq N-R,
 \label{eq:counting_main}
\end{equation}
Here $K_4=C_4/3$, and $E_{\rm err}$ sums the failure probabilities of the single-site Pauli tests, without normalization. Pairing the exceptional small clusters leaves at most one unmatched cluster, producing the offset of one. Controlling its overlap with the null input by the false-alarm probability gives the second inequality.

To relate the error count to the original task, evaluate the protocol on the allowed isotropic test channel. Its site-averaged success probability is
\begin{equation}
 P_{\rm av}(\E_\star)=(1-p_{\min})\frac FN+
 p_{\min}\left(1-\frac{E_{\rm err}}{3N}\right).
 \label{eq:average_main}
\end{equation}
Since $v\leq P_{\rm av}$ and $F\leq\alpha$, the benchmark shortfall is at least $p_{\min}E_{\rm err}/(3N)$. Equation~\eqref{eq:counting_main} then gives the lower bound in Theorem~\ref{thm:protection_main}.

For a mixed input, average the linear error and false-alarm inequalities over a pure-state decomposition with the readout held fixed. Averaging once more over retained preparation labels gives the protocol-level result. The detector has access only to the retained label; a label introduced by the mathematical decomposition supplies it with no extra information.

\subsection{The depth-two frontier and gains from larger blocks}
\label{sec:small_blocks}

A converse bounds what any allowed protocol can achieve; it need not be tight at a fixed depth. At depth two the zero-false-alarm trade-off can be solved exactly, including arbitrary global readout.

\begin{theorem}[Exact depth-two cost at zero false alarms]
\label{thm:depth_two_exact}
For all $N\geq1$ and $R\geq0$,
\begin{equation}
 V_{2,0}(N,R)=p_{\min}\left(\frac23+\frac{\min(R,N)}{3N}\right).
 \label{eq:depth_two_exact}
\end{equation}
\end{theorem}

At zero false alarms, pairing exposed probes gives no advantage over independent inputs in the stated minimax task. To see why, separate the non-null error spaces of different product factors. They are mutually orthogonal when only one site is faulty. An exposed qubit paired with a protected qubit contributes at most $p_{\min}$ to the sum of site scores. An all-exposed pair contributes at most $4p_{\min}/3$, even when its two qubits are entangled. This pair bound follows by writing the pure input in Schmidt form and optimizing the null-orthogonal measurement (Appendix~\ref{app:depth_two}).

There can be at most $R$ exposed--protected pairs. Summing the factor contributions gives Eq.~\eqref{eq:depth_two_exact}; the randomized partial-ancilla protocol attains it.

The same separation of error spaces reduces the zero-false-alarm problem at any input depth to two steps: optimize individual clusters, then allocate them across the array. Let $W^\star(m,r)$ be the largest sum of correct-site probabilities for a pure block with $m$ exposed and $r$ protected qubits, tested by the isotropic Pauli channel of strength one, with zero false alarms. This is a sum over exposed sites, not a site average. The measurement is unrestricted within the block.

\begin{proposition}[Reduction to cluster optima]
\label{prop:cluster_reduction}
In the ideal single-interrogation model,
\begin{equation}
 V_{k,0}(N,R)=\frac{p_{\min}}{N}
 \max_{\{(m_b,r_b)\}}\sum_b W^\star(m_b,r_b),
 \label{eq:cluster_reduction}
\end{equation}
where $m_b\geq1$, $r_b\geq0$, $m_b+r_b\leq k$, $\sum_bm_b=N$, and $\sum_br_b\leq R$.
\end{proposition}

The upper bound follows from orthogonal factor error spaces. To attain it, local Clifford randomization removes the unknown channel direction, and a private permutation of exposed labels equalizes the site scores. These operations preserve the input resources. The construction uses the model's freedom to configure input blocks without geometric restrictions; Appendix~\ref{app:cluster_reduction} gives the fixed-state semidefinite program and proof.

At fixed $k$, only finitely many block types remain, but each still requires optimization over a continuous input space. For $k\leq4$, only $W^\star(3,0)$, $W^\star(4,0)$, $W^\star(2,1)$, and $W^\star(3,1)$ remain to be determined; the other block values follow from the single-qubit, Bell-pair and depth-two results.

Three- and four-qubit blocks already give constructive gains. Apply a fresh, independently chosen single-qubit Clifford before each exposed channel and its inverse afterwards. Averaging this private control replaces any qubit CPTP map by the isotropic channel with the same $p(\E)$. A random permutation of exposed labels makes a general block's site-averaged score a worst-site guarantee. Neither operation increases the input depth or the number of protected ancillas.

For a graph state, let $n_{j,s}$ count the Pauli axes at exposed site $j$ that produce syndrome $s$. A syndrome decoder that reports the site with the largest such count has normalized score
\begin{equation}
 \frac{v}{p_{\min}}=\frac{1}{3N}\sum_{s\ne0}\max_j n_{j,s}.
 \label{eq:graph_frontier}
\end{equation}
This is also an upper bound for every measurement on that fixed graph input: the isotropic-test output states commute in the graph basis, so each non-null sector can contribute at most its largest site weight. The sector maxima therefore give a diagonal dual certificate: a feasible solution of the dual measurement-optimization problem that matches the achieved score and proves optimality for this fixed input.

Enumerating all labeled graphs on at most four vertices and all allowed products gives a finite, exactly certified frontier within this graph-block class (Appendix~\ref{app:graph_frontier}). With no protected ancillas, three- and four-vertex paths attain respectively
\begin{equation}
 v_{P_3}=\frac79p_{\min},\qquad
 v_{P_4}=\frac56p_{\min}.
 \label{eq:small_path_scores}
\end{equation}
For these two paths, classical randomization of the decoder suffices to equalize sites, without permuting the input labels. Each site has two Pauli axes with unique records. The remaining axis shares a record with two other sites in $P_3$, or one other site in $P_4$. Uniformly assigning a shared record gives $(2+1/3)/3=7/9$ or $(2+1/2)/3=5/6$ at every site. Local Clifford randomization is still needed for the unknown-channel guarantee. For an array tiled by identical blocks, these same scores hold whenever the block size divides $N$.

\paragraph*{Nonstabilizer constructions.}
\label{sec:nonstabilizer}

Allowing nonstabilizer inputs improves on the best graph block already at four exposed qubits. Consider the symmetric tetrahedron state~\cite{Baguette2014}, written in a convenient orientation as
\begin{equation}
 |T_4\rangle=\frac{|0000\rangle}{\sqrt3}
 +\frac{|0111\rangle+|1011\rangle+|1101\rangle+|1110\rangle}{\sqrt6}.
 \label{eq:tetra_state}
\end{equation}
Its one-qubit reductions are maximally mixed. Each two-qubit reduction is maximally mixed within the symmetric subspace, rather than in the full four-dimensional space. Thus it is not two-uniform and cannot attain the benchmark. Its error states nevertheless remain partly distinguishable even where they overlap, improving on the path's identical error states.

\begin{proposition}[Fault sensing with the tetrahedron state]
\label{prop:tetra}
Prepare $|T_4\rangle$, expose $m\in\{3,4\}$ of its qubits and protect the remainder. A square-root measurement gives, for every allowed channel at every exposed site,
\begin{equation}
 \begin{aligned}
 P_{\rm correct}&=c_m p(\E),\\
 P_{\rm wrong}&=(1-c_m)p(\E),\\
 P_{\rm none}&=1-p(\E),
 \end{aligned}
 \label{eq:tetra_response}
\end{equation}
with zero false alarms, where
\begin{equation}
 c_m=\frac{[\sqrt{m+2}+(m-1)\sqrt2]^2}{3m^2}.
 \label{eq:tetra_score}
\end{equation}
The resulting minimax score $c_m p_{\min}$ is optimal for this fixed input, without a Pauli-mixture assumption or channel twirling.
\end{proposition}

For four exposed qubits, $c_4=(2+\sqrt3)/4\simeq0.9330$, compared with $5/6$ for the best graph block. With three exposed qubits and one protected qubit, $c_3=(13+4\sqrt{10})/27\simeq0.9500$. The improvement comes from distinguishing nonorthogonal error states and permits wrong-site reports under a fault; zero false alarms alone does not forbid such reports. Appendix~\ref{app:tetra} constructs the measurement and a matching fixed-input dual bound. These constructions give lower bounds on $W^\star(4,0)$ and $W^\star(3,1)$.

This construction also bounds the best possible converse coefficient. If $C_{\rm opt}$ is the smallest constant such that $\Delta_{k,0}/p_{\min}\geq[N-R]_+/(C_{\rm opt}N)$ for all inputs of depth at most four, then
\begin{equation}
 8+4\sqrt3\leq C_{\rm opt}\leq12+6\sqrt3.
 \label{eq:constant_bracket}
\end{equation}
The lower endpoint follows by tiling an all-exposed array with tetrahedron blocks; the upper endpoint is Theorem~\ref{thm:protection_main}. The interval is approximately $[14.93,22.39]$. The cluster reduction identifies exactly which blocks determine its sharp value.

\begin{proposition}[Four-qubit characterization of the sharp coefficient]
\label{prop:sharp_constant}
The smallest universal zero-false-alarm converse coefficient at depth at most four is
\begin{equation}
 C_{\rm opt}=\max\left\{
 \frac{4}{4-W^\star(4,0)},\,
 \frac{2}{3-W^\star(3,1)}\right\}.
 \label{eq:sharp_constant_reduction}
\end{equation}
\end{proposition}

Each block has an ancilla deficit $m-r$ and a block-score loss $m-W^\star(m,r)$. Their ratio is the ancilla deficit per unit block-score loss. Because both quantities add across clusters, the largest ratio determines the universal coefficient. The existing bounds exclude every block with fewer than four qubits (Appendix~\ref{app:sharp_constant}). The remaining competition is between four exposed probes and three exposed probes sharing one protected qubit.

An intermediate bound would already simplify this problem: if $W^\star(3,1)\leq2+\sqrt3/2\simeq2.86603$, the all-exposed block alone determines $C_{\rm opt}$. Proving in addition $W^\star(4,0)=2+\sqrt3$ would close Eq.~\eqref{eq:constant_bracket} at its lower endpoint. Both statements remain open. Equation~\eqref{eq:sharp_constant_reduction} is exact at zero false alarms; it does not assert the sharp coefficient at $\alpha>0$.

A nonstabilizer block also improves on the graph construction for two exposed probes sharing one protected qubit. A partially overlapping pair of conditional sensor states gives
\begin{equation}
 W^\star(2,1)\geq\frac{41}{24}>\frac53.
 \label{eq:shared_ancilla_gain}
\end{equation}
Here local Clifford randomization and a private exchange of the two exposed roles convert a summed isotropic score into the worst-site guarantee. The explicit state, binary measurement and optimization within this family are given in Appendix~\ref{app:shared_ancilla}. Together these constructions show that graph enumeration does not close the small-cluster problem.

\par\medskip
\noindent\begin{minipage}{\columnwidth}
\captionof{table}{Zero-false-alarm scores divided by $p_{\min}$ for $N=4$, including the nonstabilizer constructions. Depth two is exact over all inputs and measurements; the other columns are constructive lower bounds from discrete block allocation, with unrestricted optimality still open. Here $a=(2+\sqrt3)/4$ and $b=(11+2\sqrt{10})/18$. Unused protected qubits are allowed.}
\label{tab:small_frontier}
\begin{ruledtabular}
\begin{tabular}{cccc}
$R$ & Depth two & Depth at most three & Depth at most four\\
\hline
0 & $2/3$ & $3/4$ & $a$\\
1 & $3/4$ & $5/6$ & $a$\\
2 & $5/6$ & $41/48$ & $b$\\
3 & $11/12$ & $89/96$ & $b$\\
4 & $1$ & $1$ & $1$
\end{tabular}
\end{ruledtabular}
\end{minipage}
\par\medskip

Table~\ref{tab:small_frontier} applies the cluster allocation to these constructions. Some intermediate budgets leave protected qubits unused; a fourth protected qubit closes the remaining gap. The same allocation with $N=1000$ gives the sufficient budgets in Fig.~\ref{fig:resource}b. Appendix~\ref{app:small_block_status} separates the analytical upper bounds from the numerical searches and the construction-allocation certificate. The targeted searches find the analytic candidates again without establishing their global optimality.

\subsection{How many exposed probes must participate}

\begin{figure*}[tbp]
\centering
\includegraphics[width=\textwidth]{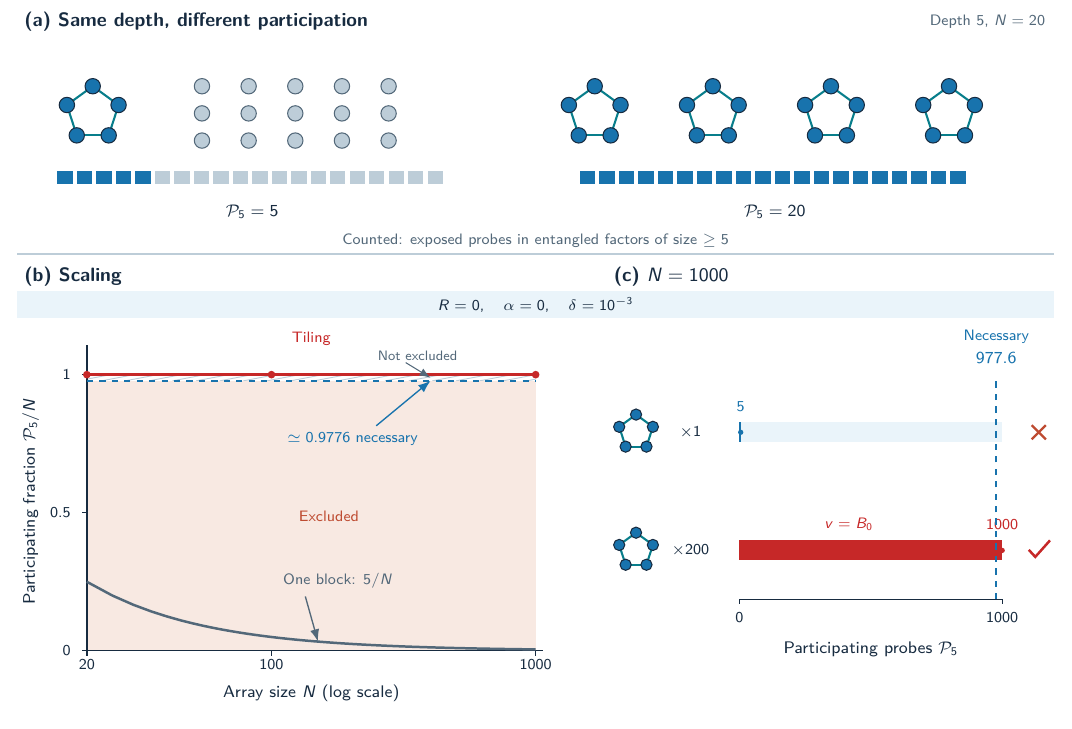}
\caption{\textbf{Approaching the benchmark without protected ancillas requires extensive participation of exposed probes.}
(a) The same depth can conceal different participation: one five-cycle and fifteen independent probes (left), or four five-cycles (right), both with $N=20$ and depth five. Blue probes belong to five-cycle factors and contribute to $\mathcal P_5$; gray probes are independent. Each segment below represents one probe. The participating fractions are $5/20$ and $20/20$.
(b) Scaling with array size. Panels (b,c) use $R=0$, $\alpha=0$ and the target $v\geq B_0-\epsilon$, with $\delta=\epsilon/p_{\min}=10^{-3}$. Theorem~\ref{thm:participation_main} requires $\mathcal P_5/N\geq1-C_4/1000\simeq0.977608$ at any depth (blue dashed line). The peach region is excluded. The hatched band is left open by the converse, which provides no achievability guarantee there. The gray curve gives the participating fraction $5/N$ for a single five-cycle; it does not plot a correct-site probability. The red line and points show full five-cycle tilings, which attain $B_0$ for $N=5m$. The horizontal axis is logarithmic.
(c) The comparison at $N=1000$. Icons and multipliers denote one or 200 five-cycles; bar lengths count participating exposed probes against the full array (pale background). A single cycle supplies five probes, below the necessary threshold $\mathcal P_5\simeq977.608$ (dashed marker), so the target is excluded (cross). A tiling of 200 cycles supplies 1000 participating probes and attains $v=B_0$ (tick).}
\label{fig:participation}
\end{figure*}

Entanglement depth specifies the largest factor size but does not count how many exposed probes participate in larger entangled factors. An array with a single five-cycle and otherwise independent probes has the same depth as an array fully tiled with five-cycles (Fig.~\ref{fig:participation}a). A useful resource description must record both how large the entangled factors are and how many probes belong to them.

The participation quantity $\mathcal P_5$ counts exposed probes in factors containing at least five qubits. For a pure input, minimize this count over exact product partitions. For a mixed input, take the infimum of the mean count over pure-state decompositions, separately for each retained preparation label (Appendix~\ref{app:participation}). Factor size includes protected ancillas; $\mathcal P_5$ counts only exposed probes.

A large entangled factor can still have indistinguishable fault locations. Participation is therefore a necessary resource count, without defining a general resource theory or a sufficient measure of sensing performance. The following bound places no upper limit on input entanglement depth.

\begin{samepage}
\begin{theorem}[Exposed-probe participation]\label{thm:participation_main}
For any labeled input ensemble and conditional global readout satisfying $F\leq\alpha$, with no restriction on input entanglement depth, the achieved worst-case score $v$ obeys
\begin{equation}
 R+\mathcal P_5+\kappa_\alpha
 \geq N-\frac{C_4N}{p_{\min}}\,[B_\alpha-v].
 \label{eq:participation}
\end{equation}
The participation measure $\mathcal P_5$ is defined formally in Eqs.~\eqref{eq:h5_main} and \eqref{eq:C5_main}.
\end{theorem}
\end{samepage}

To prove the bound, set aside factors containing at least five qubits and apply Eq.~\eqref{eq:counting_main} to the remaining small clusters. The exposed probes in the removed factors contribute to $\mathcal P_5$. Minimization over exact product partitions ensures that regrouping independent small factors cannot inflate the participation count.

At zero false alarms and without ancillas, a vanishing participating fraction leaves a finite gap from the benchmark. A single five-cycle with otherwise independent probes has $\mathcal P_5=5$. For any readout, Theorem~\ref{thm:participation_main} gives $(B_0-v)/p_{\min}\geq(1-5/N)/C_4$, so increasing the array size cannot close the gap. More generally, the same conclusion holds whenever $\mathcal P_5/N$ vanishes, even if the participating factors themselves grow.

Quantitatively, at $R=0$ and $\alpha=0$, achieving $v\geq B_0-\epsilon$ requires $\mathcal P_5/N\geq1-C_4\epsilon/p_{\min}$. Figure~\ref{fig:participation}b compares this requirement with the two inputs at $\epsilon/p_{\min}=10^{-3}$. The necessary fraction is $1-C_4/1000\simeq0.977608$, while a single five-cycle supplies only $5/N$. Full tilings have unit participation and attain $B_0$ for $5\mid N$. The hatched region is not excluded by the converse, which alone gives no guarantee of achievability there.

For $N=1000$, the necessary participation measure is at least $1000-C_4\simeq977.608$; a single cycle supplies only five (Fig.~\ref{fig:participation}c). A tiling of 200 five-cycles has $\mathcal P_5=1000$ and attains $B_0$. The value $1000-C_4$ is a necessary participation threshold. The required entanglement can remain confined to blocks of bounded size, but those blocks must involve a number of exposed probes proportional to the array size.

The participation bound also gives a resource witness from trusted, randomized single-site Pauli injections. Supplemental Material~\cite{Supplemental}, Sec.~\ref{app:witness}, derives the witness and its finite-sample and calibration corrections; the test requires a declared ancilla count and the same input and readout across fault and null trials.

\subsection{Background noise changes the interrogation cost}

The preceding benchmark assumes that unaffected sites evolve under the identity channel. Background fluctuations can produce the same records as a fault, so the false-alarm constraint limits how often the detector may act on those records. To quantify this limitation, introduce independent depolarizing background
\begin{equation}
 \mathcal D_\eta(\tau)=(1-\eta)\tau+\eta I_2/2,
 \qquad 0<\eta\leq1.
 \label{eq:background}
\end{equation}
The null channel is $\mathcal D_\eta^{\otimes N}\otimes\id_A$. Under a fault at $j$, that site's channel is $\mathcal D_\eta\circ\E$, while all other exposed sites undergo $\mathcal D_\eta$. Protected ancillas continue to undergo identity evolution.

We apply the background after the fault; Appendix~\ref{app:ordering} treats the reverse ordering for independent local background. Both orderings obey the bounds below, although an arbitrary fault channel need not commute with the background. Nonidentity backgrounds are already included in general channel-position finding~\cite{Zhuang2020}. Here we ask what a separate false-alarm budget implies for the same unknown-fault family.

\begin{proposition}[False alarms and repeated interrogation]\label{prop:noisy}
Define
\begin{equation}
 c_\eta=1+\frac{4(1-\eta)p_{\min}}{3\eta}.
 \label{eq:cp_constant}
\end{equation}
For one interrogation, every protocol, with arbitrary input entanglement, protected ancillas and readout, obeys
\begin{equation}
 v_\eta\leq\min\{B_\alpha,c_\eta\alpha/N\}.
 \label{eq:noisy_bound}
\end{equation}
If the same faulty site and channel persist for $T$ memoryless interrogations of the entire array, the bound
\begin{equation}
 v_{\eta,T}\leq\min\{1,c_\eta^T\alpha/N\}
 \label{eq:repeated_bound}
\end{equation}
holds also with hypothesis-independent adaptive controls and quantum memory between interrogations. There is one final site/no-report decision, with overall null false-alarm probability at most $\alpha$.
\end{proposition}

The key step is to compare a fault hypothesis with the null hypothesis in completely positive (CP) order: $\mathcal A\leq_{\rm CP}\mathcal B$ means that $\mathcal B-\mathcal A$ is completely positive. For the allowed isotropic test channel $\E_\star$, $\mathcal D_\eta\circ\E_\star\leq_{\rm CP}c_\eta\mathcal D_\eta$. Tensor products, ancillary extensions and intervening controls preserve this relation, so each interrogation contributes one comparison factor. The final fault output is bounded above by $c_\eta^T$ times the null output, giving Eq.~\eqref{eq:repeated_bound}. This applies established channel max-relative-entropy bounds to adaptive discrimination~\cite{Wilde2020}. Approximate unambiguous discrimination uses related but different error constraints~\cite{ZhuangAUD2020}.

The argument extends to any known background admitting a finite comparison in completely positive order. For a background channel $\mathcal B$, suppose that one allowed test channel $\E_*$ satisfies
\begin{equation}
 J(\mathcal B\circ\E_*)\leq c J(\mathcal B),\qquad c<\infty,
 \label{eq:general_cp_order}
\end{equation}
where $J$ is the Choi operator with a common normalization. Then every $T$-round protocol in the same persistent-fault model satisfies $v_T\leq\min\{1,c^T\alpha/N\}$. The least $\log c$ for a fixed pair is their channel max-relative entropy (with natural logarithms). A full-rank Choi operator $J(\mathcal B)$ guarantees a finite $c$; full-rank outputs alone do not. If Eq.~\eqref{eq:general_cp_order} fails, this converse supplies no finite comparison constant.

For any fixed $\eta>0$, strictly zero false alarms force the guaranteed correct-site probability to vanish for every finite $T$. With $0<\eta<1$, $\alpha>0$, and a fixed desired success $v_*>0$, Eq.~\eqref{eq:repeated_bound} requires
\begin{equation}
 T\geq\frac{\log(Nv_*/\alpha)}{\log c_\eta}
 \quad\text{whenever }Nv_*>\alpha.
 \label{eq:round_lower}
\end{equation}
The single-interrogation, identity-background result is therefore a singular limit of the noisy problem. The success bound holds for any number of protected ancillas: protecting the retained systems alone cannot keep the one-round success probability constant as $N$ grows.

Independent probes already achieve this order by repeating a reset--prepare--interrogate--measure cycle. In each round, every probe is independently prepared in a random Pauli eigenstate and measured in its preparation basis. The same physical qubits can be reused; only the classical records are retained between rounds. Reporting only when exactly one site reaches a count threshold suffices for $T=O(\log N)$ at fixed $\eta\in(0,1)$, $p_{\min}>0$, positive $\alpha$ and target success below one. Appendix~\ref{app:repeated} proves the guarantee over the full channel family using Hoeffding's inequality. Hence $T=\Theta(\log N)$ in this regime. Any advantage beyond this common scaling must enter the prefactor or the finite-round budget.

The conservative separation from all depth-at-most-four inputs uses a different limit, with per-probe background strength $\eta$ decreasing as $1/N$ (Supplemental Material, Sec.~\ref{app:small_noise}).

\subsection{Finite-round performance with preparation and readout costs}
\label{sec:finite_noise}

We compare repeated local monitoring, protected Bell readout and encoded readout at fixed background strength using their complete per-round record distributions. All three use the same unknown-channel family, one exposed probe per site and the same false-alarm and success targets. Each round resets and prepares independent probes, protected Bell pairs, or five-cycle blocks. The faulty site and channel persist, while preparation randomness and noise are independent between rounds. For the encoded inputs, fresh local Clifford twirling---averaging over randomly chosen single-qubit Clifford gates before the channel and their inverses afterwards---reduces the unknown channel to its isotropic representative without narrowing the allowed channel family. The controls are private and independent of the channel.

Let $P_0(s)$ be a block's complete null-syndrome distribution, including the specified circuit noise under the no-fault hypothesis, and let $u_{j,a}$ be the syndrome shift caused by $\sigma_a$ at exposed site $j$. For Pauli circuit noise and ideal twirling controls, the faulty-block distribution is exactly
\begin{equation}
 P_{j,p}(s)=(1-p)P_0(s)+\frac p3\sum_{a=X,Y,Z}P_0(s\mathbin\oplus u_{j,a}).
 \label{eq:syndrome_convolution}
\end{equation}
The complete five-cycle record has 32 outcomes. For the threshold decoder, we first compress each record to site indicators: a site's indicator is one when the record belongs to its three single-Pauli syndromes. A fault can also increase the indicator rate at another site in the same block, so a correct-site guarantee must include those competing crossings.

For blocks of $m$ exposed probes, let $q_0$ upper-bound every null indicator probability, $q_1$ lower-bound the true-site probability over the prescribed strength interval, and $q_w$ upper-bound the indicator probability of any wrong site in the faulty block. Equation~\eqref{eq:syndrome_convolution} is affine in $p$, so the required extrema occur at $p_{\min}$ or $p_{\max}$. Set
\begin{equation}
 U_T(q,h)=\sum_{r=h}^{T}\binom{T}{r}q^r(1-q)^{T-r}.
 \label{eq:binomial_tail}
\end{equation}
Report a site only if it is the unique site whose count reaches $h$. Each count is binomial, although counts within one block can be correlated. Union bounds give
\begin{align}
 F&\leq N U_T(q_0,h),\nonumber\\
 1-v&\leq1-U_T(q_1,h)+(N-m)U_T(q_0,h)\nonumber\\
 &\hspace{24mm}+(m-1)U_T(q_w,h).
 \label{eq:finite_round_certificate}
\end{align}
The bounds require no independence between different sites' indicators. The three failure terms respectively bound a missed true-site crossing, a competing crossing outside the faulty block, and one within it.

Without circuit noise, independent probes have $q_0=\eta/2$ and $q_1=q_0+2(1-\eta)p_{\min}/3$; Bell pairs have $q_0=3\eta/4$ and $q_1=q_0+(1-\eta)p_{\min}$. Both have $m=1$. Chernoff bounds for these Bernoulli counts, with $q_0<\tau<q_1$, give
\begin{align}
 F&\leq N e^{-T d(\tau\Vert q_0)},\nonumber\\
 1-v&\leq e^{-T d(\tau\Vert q_1)}+(N-1)e^{-T d(\tau\Vert q_0)},
 \label{eq:kl_rounds}
\end{align}
where $d$ is the relative entropy between Bernoulli distributions, $d(x\Vert y)=x\log(x/y)+(1-x)\log[(1-x)/(1-y)]$. This rule compares the empirical rate with $\tau$. For a five-cycle, add the within-block term $(m-1)e^{-T d(\tau\Vert q_w)}$ and replace $N-1$ by $N-m$, provided $\max(q_0,q_w)<\tau<q_1$.

These exponents improve the quadratic Hoeffding estimate for rare background events. The numerical comparison evaluates the binomial sums themselves and optimizes over integer thresholds.

\begin{figure*}[t]
\centering
\includegraphics[width=\textwidth]{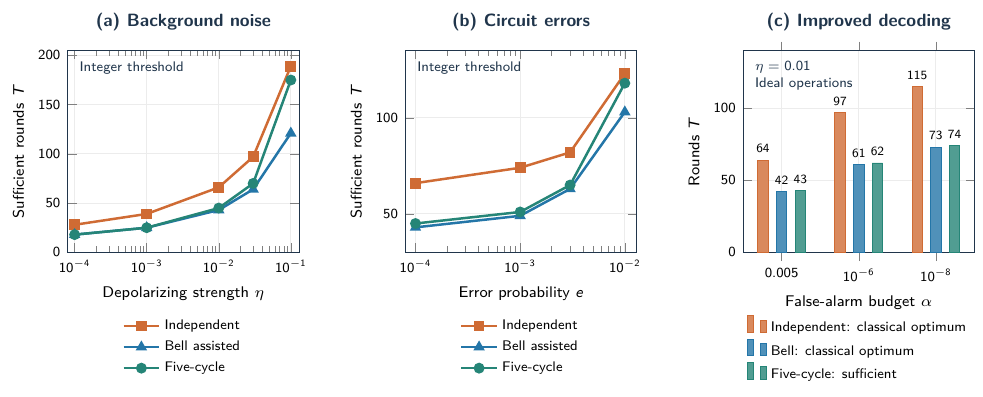}
\caption{\textbf{Sufficient interrogation budgets for specified readouts.}
All panels use $N=1000$, $p\in[0.2,1]$ and target $v\geq0.9$.
(a,b) Integer-threshold decoding with $F\leq0.005$. Each point is the smallest round count certified by Eq.~\eqref{eq:finite_round_certificate} among integer thresholds; it is a sufficient budget, not an unrestricted optimum. (a) Depolarizing background with ideal operations. (b) At $\eta=0.01$, initialization and measurement bit errors each have probability $e$, and every two-qubit gate is followed by a uniformly random nonidentity two-qubit Pauli with total probability $e$. One-qubit controls and protected storage are ideal.
(c) Improved classical decoding at $\eta=0.01$ with ideal operations, for three array false-alarm budgets. Independent and Bell values are classical optima for the specified fixed Bernoulli readouts. Five-cycle values are sufficient full-syndrome likelihood guarantees. At $\alpha=0.005$ the budgets are 64, 42 and 43: five-cycle readout uses fewer rounds than independent probes and one more than the ideal Bell readout, with no protected input qubits. These improvements do not recalibrate the noisy-gate curves in (b). The four-probe construction comparison is retained in Supplemental Material, Fig.~\ref{fig:small_frontiers}.}
\label{fig:finite_tradeoffs}
\end{figure*}

For these threshold decoders, five-cycle readout reduces the round budget relative to independent probes and approaches Bell performance without protected input qubits. At $\eta=0.01$, $N=1000$, $p\in[0.2,1]$, $\alpha=0.005$ and target $v\geq0.9$, the sufficient budgets are 66 rounds for independent probes, 43 for Bell pairs and 45 for five-cycles.

Preparation and decoding noise increase these budgets. With the noise model of Fig.~\ref{fig:finite_tradeoffs}b and $e=10^{-3}$, they become 74, 49 and 51 rounds; at $e=10^{-2}$ they are 123, 103 and 118. These increases quantify the noise cost of the specified preparation and decoding circuits.

To count circuit resources, prepare a five-cycle with five controlled-$Z$ gates and decode by their inverse followed by local measurements. This uses ten two-qubit gates and five measured qubits per block per round, with no temporary readout ancilla. A Bell pair uses one entangling gate for preparation and one for decoding, and measures two qubits.

The Bell and five-cycle circuits therefore use $2NT$ two-qubit gates in these circuits. Five-cycle readout measures $NT$ qubits; Bell readout measures $2NT$. Independent probes use no two-qubit gates and measure $NT$ qubits. Bell probing additionally retains $N$ protected qubits throughout each interrogation. Local Clifford controls, one-qubit gates and their timing must be budgeted in an implementation; they are ideal in this noise calculation. The circuits implement direct decoding without fault-tolerant syndrome extraction.

Noise in the retained ancillas weakens Bell-assisted readout. The released data include depolarization $\eta_A=0.01$ of the Bell partners during exposure. With $\eta=0.01$ and otherwise ideal operations, the Bell sufficient budget becomes 54 rounds. This memory-noise calculation extends beyond the ideal protected-ancilla resource model. It and Fig.~\ref{fig:finite_tradeoffs} compare the stated repeated-readout protocols; the earlier converses allow arbitrary global measurements and constrain a larger class of independent-input strategies.

We next improve the classical decoder while keeping these quantum record distributions fixed.

\paragraph*{Decoding complete records.}

The round counts above depend on the final classical decision. Integer thresholds leave part of the false-alarm budget unused, and discarding all outcomes with competing crossings loses location information. For independent-site Bernoulli readouts, the optimal decoder accounts for both effects.

\begin{proposition}[Optimal classical decoding of Bernoulli records]
\label{prop:optimal_count}
Suppose each of $N$ sites produces $T$ independent Bernoulli records. Under the null each record equals one with probability $q_0$; under a fault at site $j$, that probability is at least $q_1>q_0$, while the other sites retain $q_0$. Records at different sites are independent. For this fixed readout, an optimal worst-site decoder selects a site with the largest count and splits ties uniformly. It reports only when that maximum exceeds a threshold, with possible randomization at the boundary. Calibrating the rule under the null attains the constraint $F\leq\alpha$.
\end{proposition}

At the weakest fault, the site likelihood ratio increases with its count. Maximizing the average correct-report probability at a fixed null-report cost therefore selects the largest count, with a common rejection threshold. Permutation symmetry makes the site-average and worst-site values equal. The correct-report probability increases with the faulty site's count, so the weakest fault is also the least favorable. Appendix~\ref{app:optimal_decoding} gives the exact finite-sample expression, including ties. An allowed isotropic test channel makes this decoder classically optimal for the fixed twirled independent-probe and Bell readouts. The quantum inputs and measurements remain fixed in this optimization.

For a five-cycle, retain the full syndrome $s$ and form each site's matched log-likelihood score
\begin{equation}
 \ell_j(s)=\log\frac{P_{j,p_{\min}}(s)}{P_0(s)}.
 \label{eq:site_likelihood}
\end{equation}
We sum quantized per-round scores, placing each round's $\ell_j$ in bins of $1/32$ nat. Report only a unique site that crosses the resulting threshold, randomizing a boundary score to use the false-alarm budget.

A guarantee uniform in fault strength requires separate bounds on missed and wrong-site crossings. The true-site score increases stochastically with $p$, because it is an increasing function of the one-round likelihood ratio. Thus $p_{\min}$ controls misses for this rule. A stochastic envelope of the two endpoint distributions bounds each wrong-site crossing throughout the channel interval. The union-bound certificate follows the logic of Eq.~\eqref{eq:finite_round_certificate}, now using the complete record. Appendix~\ref{app:likelihood_envelope} proves this sufficient guarantee. The joint indicator calculation remains available in Supplemental Material, Sec.~\ref{app:joint}.

\par\medskip
\noindent\begin{minipage}{\columnwidth}
\captionof{table}{Round budgets at $N=1000$, $\eta=0.01$ and target $v\geq0.9$, with ideal gates and controls. Five-cycle entries are sufficient full-record likelihood budgets; the other entries are classical optima for the specified repeated Bernoulli readouts. The noisy Bell-memory row uses $\eta_A=0.01$. Using two independent sensors per module requires $2N$ exposed sensors and keeps both bits. Known damping changes the channel and background families. The quantum inputs are fixed in every row.}
\label{tab:rare_alarm}
\begin{ruledtabular}
\begin{tabular}{lrrr}
Array false-alarm budget $\alpha$ & $0.005$ & $10^{-6}$ & $10^{-8}$\\
\hline
Independent & 64 & 97 & 115\\
Bell, ideal memory & 42 & 61 & 73\\
Five-cycle (sufficient) & 43 & 62 & 74\\
Bell, noisy memory & 52 & 80 & 93\\
Two local independent sensors & 32 & 49 & 58\\
Known amplitude damping & 21 & 31 & 37
\end{tabular}
\end{ruledtabular}
\end{minipage}
\par\medskip

Changing only the classical processing reduces the gap between the round budgets (Table~\ref{tab:rare_alarm}). At $\alpha=0.005$, boundary randomization alone reduces the independent, Bell and five-cycle sufficient budgets from $66,43,45$ to $65,42,44$. Optimal largest-count decoding further reduces the independent budget to 64, while the complete five-cycle record certifies 43.

At the two rarer false-alarm targets, the five-cycle guarantees are 62 and 74 rounds, compared with 97 and 115 for independent readout and 61 and 73 for ideal Bell readout. Thus the encoded protocol improves on the specified independent readout while remaining one round above Bell readout without protected storage. The five-cycle decoder is not known to be optimal. These comparisons also show how classical processing changes the relative round budgets for the same quantum records.

The false-alarm probability refers to one completed array decision. At $f$ completed decisions per second, the expected number of null false alarms per hour is at most $3600f\alpha$, without requiring independence between decisions. For illustration, one decision per second and $\alpha=0.005$ permit 18 per hour; $\alpha=10^{-6}$ permits 0.0036. These examples assume a decision cadence; they make no device timing prediction. An event prior is additionally needed to infer the reliability of a reported event or location.

\paragraph*{Resource and mechanism dependence.}

A shorter round budget can be purchased with additional local sensors. Two independent sensors per module provide a useful baseline if both experience product fault channels of strength at least $p_{\min}$, with independent backgrounds and private controls. Keeping both measured bits makes the total count over $T$ rounds binomial with $2T$ trials, so Proposition~\ref{prop:optimal_count} applies. At $\alpha=0.005$, 32 rounds use $2N$ exposed sensors and $64N$ channel queries. The five-cycle guarantee uses $N$ exposed sensors and $43N$ queries, together with $86N$ two-qubit gates. The shorter local protocol exchanges additional sensors and queries for fewer parallel rounds. Its shared-footprint and product-channel assumptions must be checked for a physical module.

Excited-state monitoring of relaxation bursts~\cite{McEwen2022} motivates a separate known-mechanism comparison. Here both the fault family and the background model change; this is not a reconstruction of that experiment. For known amplitude damping, a process infidelity relative to identity of 0.2 corresponds to damping probability $\gamma\simeq0.37771$. Matching the background process infidelity $3\eta/4=0.0075$ gives damping probability $\gamma_0\simeq0.014972$. Excited-state probes then have flag probabilities $q_0=\gamma_0$ and $q_1=\gamma_0+(1-\gamma_0)\gamma$. Their optimal fixed-readout budget is 21 rounds at $\alpha=0.005$, compared with 22 for the original threshold. Uncertainty in the damping rate alone does not require an encoded detector; the full-family problem also leaves the channel mechanism and direction unspecified.

The five-cycle edges connect sensors in different modules, whereas a protected Bell partner may be local. The resulting gate cost can affect the round budget. Under the original integer-threshold rule, varying only the uniformly random two-qubit Pauli error gives five-cycle budgets of 59 rounds at $e=0.005$, 66 at $e=0.006$--$0.007$, and 73 at $e=0.008$.

The 66-round crossing with the original independent rule applies to this circuit, noise model and decoder. Average gate infidelity is $4e/5$ for this channel, so $e=0.006$ corresponds to 99.52\% average fidelity; this value describes the scan and imposes no necessary fidelity requirement on each physical gate. The decoder improvements in Table~\ref{tab:rare_alarm} assume ideal gates and do not recalibrate that crossing. The separate memory and gate scans leave the joint region of architectural advantage undetermined.

These comparisons identify the relevant choices. Protected Bell readout retains $N$ protected input qubits; the five-cycle replaces that storage with correlations and gates between exposed sensors; using two independent sensors per module requires more sensors and queries; mechanism-specific probing uses stronger calibration assumptions. The preferred architecture depends on the available storage, sensors and intermodule gates. Geometry, leakage, idle errors and gate durations require a device model beyond the distributions used here.

\subsection{Information retained by a fixed readout}
\label{sec:record_information}

How close are these round budgets to the information limit of a fixed readout? A complementary lower bound depends only on the information in the specified per-round records. Fix the input, controls and quantum readout, and let $P_0$ and $P_j$ be the complete classical record distributions under the null and an allowed test fault at site $j$. Retained control labels belong to the record. Suppose rounds are independent and identically distributed and $D(P_j\Vert P_0)=D_1<\infty$ for every site, where logarithms are natural.

\begin{proposition}[Fixed-readout interrogation bound]
\label{prop:record_bound}
Any final classical decoder of $T$ such records with $F\leq\alpha$ and worst-site success at least $v_*>\alpha/N$ satisfies
\begin{equation}
 T D_1\geq d(v_*\Vert\alpha/N).
 \label{eq:record_bound}
\end{equation}
If also $D(P_0\Vert P_j)=D_0<\infty$ is site independent, then $T D_0\geq d(\alpha/N\Vert v_*)$.
\end{proposition}

To prove the first inequality, choose a site whose null report probability is at most $\alpha/N$. Data processing from the entire record to the event ``report this site'' gives $T D_1\geq d(v_j\Vert F_j)\geq d(v_*\Vert\alpha/N)$. Interchanging the two hypotheses proves the second. The argument applies to every classical decision rule after the fixed readout; it does not optimize the input, quantum measurement or adaptive quantum controls. For a minimax task, one allowed test channel suffices to constrain a uniform guarantee.

At $p=0.2$ and $\eta=0.01$, use the isotropic test channel for the full-family protocols. Its statistics are independent of the private Clifford labels, so including those labels does not change $D_1$. Independent probes, Bell readout and complete five-cycle syndromes give respectively
\begin{equation}
 D_1=0.330715,\quad0.503529,\quad0.495145
 \quad\text{nats per round}.
 \label{eq:record_information_values}
\end{equation}
For $N=1000$, $\alpha=0.005$ and $v_*=0.9$, Eq.~\eqref{eq:record_bound} requires at least 33, 22 and 22 rounds. The optimal classical budgets for the independent and Bell readouts are 64 and 42; the five-cycle sufficient budget is 43. The divergence bound is loose even where the classical optimum is known, and its equal 22-round lower bounds leave the relative Bell and five-cycle optima unresolved.

Record resolution matters in comparisons. Keeping both bits from two independent local sensors gives $D_1=2(0.330715)=0.661431$, whereas retaining only their OR flag, as in the original indicator baseline, gives $0.615478$. A single true-site flag from the five-cycle has divergence $0.487071$, below that of the complete 32-symbol syndrome. Under its different channel assumptions, the known-damping binary record gives $0.967970$. Relative entropy therefore separates information lost in readout or coarse graining from losses introduced by a particular final decision rule.

\section{Discussion}

Independent probes and protected Bell pairs expose two distinct resource costs of fault sensing. For one interrogation of an unknown local channel without background noise, independent probes achieve a zero-false-alarm score of $2p_{\min}/3$; Bell pairs attain the optimum $p_{\min}$ but retain one protected qubit per exposed probe. Suitable code states attain the same optimum with every qubit exposed. We have characterized exactly the required structure: the input support and all single-site Pauli error supports must be mutually orthogonal. For pure inputs, this is equivalent to maximally mixed one- and two-qubit reductions on exposed probes. Cycle states and products of suitable code states realize this structure in blocks whose size need not grow with the array.

The resource bounds quantify the cost of falling short of this structure. Without protected ancillas, pairs offer no improvement over independent probes, while three- and four-qubit constructions give intermediate scores. The four-qubit tetrahedron state has a matching readout and fixed-input dual certificate, but its optimality over all four-qubit inputs is not established. Depth alone also misses an extensive requirement: approaching the optimum without protected ancillas requires a number of probes proportional to the array size to participate in entangled factors of at least five qubits. Trusted Pauli tests can certify this participation, with explicit statistical and calibration penalties.

Repeated independent-probe measurements remain a strong alternative in background noise. For a persistent site and channel under memoryless depolarization of fixed strength $0<\eta<1$, they attain the optimal $\Theta(\log N)$ interrogation order at fixed success between zero and one and positive false-alarm budget. The finite-round comparison then concerns storage, queries and gate costs. In the 1000-site comparison at $\alpha=0.005$ (Table~\ref{tab:rare_alarm}), the specified independent and ideal Bell readouts require 64 and 42 rounds, while complete five-cycle records certify 43 with no protected qubits. The other two false-alarm budgets show the same one-round gap to Bell readout. Independent probes avoid entangling gates; additional local sensors can reduce parallel rounds, and knowledge of a relaxation mechanism can favor excited-state probes. The unknown-channel guarantees therefore establish a resource advantage in the stated task, while the preferred hardware protocol depends on these additional costs and information.

Physical use requires the stated channel and timing assumptions. A module must induce an input-independent local channel, with faults independent of private preparation randomness. Even ideal gates can turn a fault during decoding into a report at a different site (Appendix~\ref{app:timing}). Two-site Pauli faults can likewise produce a recognized syndrome assigned to a third, unaffected site (Supplemental Material, Sec.~\ref{sm:multiple}). Thus zero false alarms under the null hypothesis does not exclude wrong-site reports outside the single-site fault family. Extending the resource laws to continuous event onset or multiple affected modules requires a specified fault family, timing model and report convention; interrogation counts alone do not determine detection latency.

Two optimizations remain. In the identity-background problem, the unrestricted small-block optima determine the sharp resource coefficient through the four-qubit reduction. Under fixed nonzero background noise, our exact classical decoders apply to specified readouts, and the encoded round counts are sufficient guarantees. Optimizing quantum inputs, measurements and adaptation across rounds would establish whether protected storage can still be replaced at the best finite interrogation cost. The present results provide analytic resource bounds and explicit protocols against which those improvements can be measured.

\begin{acknowledgments}
This work was supported by the Key-Area Research and Development Program of Guangdong Province (Grant No.~2018B030326001), the National Natural Science Foundation of China (U1801661), the Guangdong Innovative and Entrepreneurial Research Team Program (2016ZT06D348), the Guangdong Provincial Key Laboratory (Grant No.~2019B121203002), the Natural Science Foundation of Guangdong Province (2017B030308003), the Science, Technology and Innovation Commission of Shenzhen Municipality (JCYJ20170412152620376 and KYTDPT20181011104202253), the NSF of Beijing (Grant No.~Z190012), the Shenzhen Science and Technology Program (KQTD20200820113010023), and the National Natural Science Foundation of China (Grant Nos.~12404566 and 1230041239).
DeepSeek Harness and WorkBuddy assisted with literature comparison, code and figure development, and manuscript editing under the author's direction. The author verified the resulting content and takes full responsibility for the results and the manuscript.

\end{acknowledgments}

\par\medskip
\section*{Competing interests}
The author declares no competing interests.

\section*{Data and code availability}
Code and computational data supporting this work are available from the corresponding author upon reasonable request by email (dengxiuhao@iqasz.cn). These computational data support the stated models and contain no experimental measurements.

\appendix
\section{Structural conditions and sensing protocols}
\label{app:resource}
\label{app:extensions}

\subsection{Proof of the support criterion}
\label{app:support_proof}

The key step is to examine equality in the isotropic-channel bound. For the fixed input and readout, define
$e_{ja}=1-\operatorname{Tr}(M_jP_{j,a}\rho P_{j,a})\geq0$.
At $F=0$, the site-averaged success probability for the allowed isotropic channel is
$p_{\min}[1-\sum_{ja}e_{ja}/(3N)]$. Attaining $v=p_{\min}$ therefore forces every $e_{ja}=0$. Positivity, together with the strict positivity of $\rho$ on its support, then gives
\begin{equation}
 M_jQ=0,\qquad (I-M_j)P_{j,a}Q=0 .
 \label{eq:support_saturation}
\end{equation}
At $F=0$, the first identity holds for each $j$ separately and separates the ranges of $Q$ and $P_{j,a}Q$. For different sites, $j\ne l$, the inequality $M_j+M_l\leq I$ also makes the ranges of $P_{j,a}Q$ and $P_{l,b}Q$ orthogonal, so $QP_{j,a}P_{l,b}Q=0$.

At a single site, distinct Pauli axes satisfy $P_{j,a}P_{j,b}=\pm iP_{j,c}$; compression to $Q$ vanishes because the error support is orthogonal to the null support. The diagonal products are identity. These observations establish Eq.~\eqref{eq:support}.

For sufficiency, Eq.~\eqref{eq:support} makes $P_\mu QP_\mu^\dagger$ mutually orthogonal projectors. Substituting the local Kraus expansion into the readout of Eq.~\eqref{eq:general_readout}, with $\rho=Q\rho Q$, yields Eq.~\eqref{eq:general_response}. Trace preservation gives $\sum_\ell(|a_\ell|^2+\sum_a|b_{\ell a}|^2)=1$, while the definition of fault strength gives $p(\E)=\sum_{\ell,a}|b_{\ell a}|^2$. The readout attains the benchmark.

For labeled preparations, a nonnegative false-alarm or error term can average to zero only if it vanishes for every label of positive probability. The argument therefore applies to each conditional support and its own POVM. With continuous labels, it applies almost everywhere.

For a pure input with every qubit exposed, the compressed operators reduce to scalar Pauli expectations. The one- and two-site expectations vanish precisely when the corresponding reduced states are maximally mixed.

For a general input of rank $K$, a separate dimension count applies. The $1+3N$ orthogonal supports each have dimension $K$, which proves Eq.~\eqref{eq:packing}. This is a necessary dimension bound; code existence requires additional conditions.

\subsection{Calibrated readout and approximate input structure}
\label{app:approximate}

With a readout calibrated to a known pure preparation, exact orthogonality is not needed for a useful correct-site guarantee. For a block input $|\psi_b\rangle$, form the column matrix $V_b=(P_\mu|\psi_b\rangle)_\mu$ and its \emph{complex} Gram matrix $H_b=V_b^\dagger V_b$. Unlike the squared-overlap plot in Fig.~\ref{fig:distinguishability}a, this matrix retains the phases of the overlaps.

Suppose $H_b$ is nonsingular and put $\lambda_b=\lambda_{\min}(H_b)$. The columns of $W_b=V_bH_b^{-1}$ are reciprocal vectors, a standard tool of unambiguous state discrimination~\cite{Chefles1998}. They define a physical block POVM with site effects
\begin{equation}
 M_j^{(b)}=\lambda_b\sum_a|w_{j,a}\rangle\langle w_{j,a}|.
 \label{eq:reciprocal}
\end{equation}
The residual effect returns no report. A single local channel in the block is then assigned to the correct site with probability $\lambda_b p(\E)$, with zero false-alarm and wrong-site probabilities. Preparing and reading out the blocks independently therefore guarantees
\begin{equation}
 v_0\geq p_{\min}\min_b\lambda_b.
 \label{eq:gram_robust}
\end{equation}
If $\|H_b-I\|_{\rm op}\leq\varepsilon<1$ for every block, the relative shortfall is at most $\varepsilon$, regardless of the number of blocks. Only the affected block incurs this penalty. The guarantee assumes a known pure preparation and a readout calibrated to it. Unknown mixed-state preparation noise with a fixed decoder requires a different analysis, as does the background noise in Sec.~\ref{sec:finite_noise}.

We can also ask what near-optimal performance requires of the local reduced states. Let $\delta=(B_\alpha-v)/p_{\min}$ and let $\rho^{(c)}_{jl}$ be the two-exposed-probe reduction of the conditional input. For $N\geq2$, the overlap inequalities used in our converse imply
\begin{equation}
 \begin{aligned}
 &\frac{\sum_c\lambda_c\sum_{j<l}\|\rho^{(c)}_{jl}-I_4/4\|_2^2}
 {\binom N2}
 \\
 &\leq\left(3+\frac{3h_N}{2(N-1)}\right)\delta+\frac{3\alpha}{N}.
 \end{aligned}
 \label{eq:approx_necessary}
\end{equation}
Here $\|\cdot\|_2$ is the Hilbert--Schmidt norm. The bound limits the mean squared deviation of the two-probe marginals from maximal mixing in terms of the performance shortfall and the false-alarm budget. This necessary condition gives no worst-pair guarantee and, for mixed inputs, does not replace the support criterion. Thus Eqs.~\eqref{eq:support}, \eqref{eq:gram_robust}, and \eqref{eq:approx_necessary} provide, respectively, an exact support criterion, a calibrated sufficient guarantee and an averaged necessary condition.

\paragraph*{Proofs.}

Reciprocal vectors distinguish the error states while allowing an inconclusive outcome. For a block with $H=V^\dagger V>0$, let $W=VH^{-1}$. Then $W^\dagger V=I$ and $W^\dagger W=H^{-1}$. The nonzero eigenvalues of $WW^\dagger$ are those of $H^{-1}$, so $\lambda_{\min}(H)WW^\dagger\leq I$. Omitting the reciprocal null vector preserves the inequality, proving that Eq.~\eqref{eq:reciprocal} defines a POVM.

We next evaluate its response to an arbitrary local channel. For a Kraus output $K_\ell^{(j)}|\psi\rangle=V c_\ell$, reciprocity gives $\langle w_{l,a}|K_\ell^{(j)}|\psi\rangle=\delta_{jl}b_{\ell a}$. Summing the outcome probabilities yields $\lambda_{\min}(H)p(\E)$ for the correct site and zero for every wrong site and for any report under the null input. In a product of blocks, every unaffected block returns no report. Only the affected block incurs a success penalty, establishing Eq.~\eqref{eq:gram_robust}.

To derive the necessary condition, first take a pure input component while keeping the actual conditional POVM fixed. Let $A_1$ and $A_2$ be the sums of squared expectations of weight-one and weight-two Paulis on exposed sites. The null-overlap inequality and Eq.~\eqref{eq:frame_bound_main} give
\begin{equation}
 A_1\leq2E_{\rm err}+6F,\qquad
 A_2\leq h_NE_{\rm err}.
 \label{eq:overlap_sums_ext}
\end{equation}
Expanding each two-probe reduction in the orthogonal Pauli basis gives
\begin{equation}
 \sum_{j<l}\|\rho_{jl}-I_4/4\|_2^2
 =\frac{(N-1)A_1+A_2}{4}.
 \label{eq:marginal_sum}
\end{equation}
Dividing by $\binom N2$ and using Eq.~\eqref{eq:overlap_sums_ext} bounds this average by $[2(N-1)+h_N]E_{\rm err}/[2N(N-1)]+3F/N$. Convexity of the squared Hilbert--Schmidt norm extends the bound to each mixed conditional state. Linearity preserves the averaged values of $E_{\rm err}$ and $F$. Equation~\eqref{eq:average_main} finally gives $E_{\rm err}\leq3N(B_\alpha-v)/p_{\min}$ and $F\leq\alpha$, completing the proof of Eq.~\eqref{eq:approx_necessary}.

\subsection{Channel responses and the partial-ancilla protocol}
\label{app:responses}

The Bell-ancilla and two-uniform sensors have the same response because both resolve the identity and single-site Pauli sectors. To calculate that response, expand each Kraus operator as
\begin{equation}
 K_\ell=a_\ell I+\sum_{b=X,Y,Z}b_{\ell b}\sigma_b .
 \label{eq:kraus_main}
\end{equation}
Taking the trace of the trace-preservation condition gives $\sum_\ell(|a_\ell|^2+\sum_b|b_{\ell b}|^2)=1$. Equation~\eqref{eq:strength} therefore gives $p(\E)=\sum_{\ell,b}|b_{\ell b}|^2$.

A Bell measurement on a probe--ancilla pair, or the two-uniform sector readout in Eq.~\eqref{eq:general_readout}, distinguishes the identity contribution from the local-error contributions. Stabilizer measurements implement this readout for the cycles. The null outcome has probability $1-p$, and all non-null probabilities sum to $p$. Orthogonality gives these probabilities on the original channel output, including when its Kraus operators contain coherent Pauli amplitudes.

For an independent probe, choose uniformly among the six Pauli eigenstates and measure in the corresponding basis. The mean probability of detecting a change is
\begin{equation}
 \frac16\sum_{\psi\in\{\pm x,\pm y,\pm z\}}
 \left[1-\sum_\ell|\langle\psi|K_\ell|\psi\rangle|^2\right]
 =\frac23p(\E).
 \label{eq:click_main}
\end{equation}
Averaging over opposite eigenvalue signs cancels the linear cross terms. Averaging over the three axes then sums the squared Pauli coefficients, yielding the qubit average-fidelity identity~\cite{Nielsen2002}.

For the matching converse, apply the isotropic test channel to a pure product input. The output is a mixture of the unchanged input, with probability $1-2p_{\min}/3$, and the state in which the affected qubit has been flipped to its orthogonal state, with probability $2p_{\min}/3$. Only the latter component can reveal the location. The site-averaging argument used for the Bell benchmark therefore gives Eq.~\eqref{eq:independent}, even for global readout. Averaging over pure components and retained preparation labels extends the bound to every allowed independent-probe protocol.

The partial-ancilla protocol interpolates between the two responses. For $R<N$, choose uniformly which $R$ exposed sites receive Bell ancillas, and prepare independent probes at the remaining sites. Choose a default report site uniformly among those without ancillas and retain all preparation choices. A detected change identifies its site. If all records are null, report the default site with probability $\alpha$. The false-alarm probability is then exactly $\alpha$, and every channel and fault location has correct-site probability
\begin{equation}
 P_{\rm con}(p)=
 \frac{pR+\frac23p(N-R)+\alpha(1-\frac23p)}{N}.
 \label{eq:construct_main}
\end{equation}
The faulty site receives an ancilla with probability $R/N$ and is chosen as the default with probability $1/N$. A site cannot receive an ancilla and be the default report site in the same preparation. The resulting response is uniform over sites and nondecreasing in $p$. Evaluate it at $p_{\min}$ and subtract from $B_\alpha$ to obtain the constructive shortfall bound in Eq.~\eqref{eq:law}. When $R\geq N$, use $N$ Bell pairs and leave the surplus ancillas unused.

\subsection{Exact attainment and finite private randomization}
\label{app:attainment}

To determine exact attainment, inspect where the converse can be saturated. Suppose a protocol of input depth $k\in\{2,3,4\}$ attains $v=B_\alpha$. Equation~\eqref{eq:average_main} forces $E_{\rm err}=0$. Nonnegativity of every term means that each pure component with positive weight must identify the site of every Pauli test state perfectly. The small-cluster constraints then exclude all positive-deficit clusters except $(1,0)$ and $(2,1)$, and exclude any pair of exceptional clusters. For $R<N$, exactly one exceptional cluster must remain, giving $R=N-1$.

A $(1,0)$ cluster is a pure exposed qubit. For the remaining $(2,1)$ case, zero error implies $u=0$, so the exposed reduction is
\begin{equation}
 \rho_E=\tfrac14\left(I\otimes I+
 \bm a\cdot\bm\sigma\otimes I+I\otimes\bm b\cdot\bm\sigma\right).
 \label{eq:exceptional_reduction_main}
\end{equation}
The eigenvalues of this reduction are $(1\pm x\pm y)/4$, where $x=|\bm a|$ and $y=|\bm b|$. Positivity and rank at most two force $\{x,y\}=\{1,0\}$, so this cluster also contains a pure exposed qubit at some site $j$.

In either exceptional case, let $\bm n$ be the unit Bloch vector of the pure exposed qubit. Perfect site identification for Pauli test states gives $M_j\sigma_{ja}\ket\psi=\sigma_{ja}\ket\psi$ for each axis. But $(\bm n\cdot\bm\sigma_j)\ket\psi=\ket\psi$, so the same report effect satisfies $M_j\ket\psi=\ket\psi$. Thus $F=1$: the detector must report a site even on the undisturbed input. Averaging over pure components and retained labels preserves this value, excluding $R<N$ whenever $\alpha<1$. Bell pairs attain the benchmark at $R=N$.

At $\alpha=1$, choose the sole unpaired site uniformly at random, retain its label, and give it the default report when all $N-1$ Bell records are null. A fixed actual site is unpaired with probability $1/N$ and otherwise has correct-report probability $p$, yielding $1/N+(N-1)p/N=B_1$. Fixing the unpaired site in advance would not attain this worst-site score. This establishes Corollary~\ref{cor:exact_main}.

One step remains: excluding equality for each protocol does not by itself exclude a sequence approaching equality. We close this possibility by showing that an optimal protocol exists.
\begin{proposition}[Finite realization of the optimum]\label{prop:finite_main}
For fixed $N,R,k$ and $0\leq\alpha\leq1$ in the stated channel family, $V_{k,\alpha}(N,R)$ is attained. At most $16N^2+2$ retained preparation labels suffice.
\end{proposition}

Fix a partition into clusters of at most $k$ qubits. Its pure product states form a compact set. There are finitely many partitions, so their union and its convex hull are compact as well. This convex hull is the allowed set of conditional inputs. The set of POVMs is also compact.

For each conditional input/readout pair, record the null reporting probability. For each of the $N$ possible fault sites, also record the probabilities of all $N$ site reports. Each fault response is linear in the qubit Choi matrix. The real coefficients of these responses, together with the null probability, form a compact set of dimension at most $1+16N^2$.

Private randomization forms the convex hull of the response set. Carath\'eodory's theorem bounds the required number of preparation labels by $16N^2+2$. The closed constraint $F\leq\alpha$ preserves compactness. Minimizing over the compact channel family and the finite site set gives a continuous objective, whose maximum is therefore attained. The label count proves existence; it does not estimate the cost of implementing the randomization.

\section{Resource bounds and small-block optimization}
\label{app:resource_bounds}

\subsection{Small-cluster constraints and the converse bound}
\label{app:converse}

The converse has two steps. Small clusters force overlaps, and the readout constraints turn those overlaps into errors. For a pure cluster containing $m$ exposed and $r$ protected qubits, let $s$ and $u$ be the sums of squared expectation values of Pauli operators supported on one and two exposed qubits, respectively. Table~\ref{tab:small_clusters} lists all cluster types with $m+r\leq4$ and positive ancilla deficit $d=m-r$.

\begin{center}\begin{minipage}{\columnwidth}
\centering
\captionof{table}{\textbf{Overlaps forced by pure input clusters of at most four qubits.}
A cluster contains $m$ exposed probes and $r$ protected ancillas, with deficit $d=m-r$. The quantities $s$ and $u$ sum squared expectation values of Pauli operators supported on one and two exposed probes. For $(m,r)=(3,1)$, $t$ is the protected qubit's squared Bloch-vector length. The constraints require overlap between states produced by faults at different sites or between an error state and the undisturbed input.}
\label{tab:small_clusters}
\begin{tabular}{ccl}
\toprule
$(m,r)$ & $d$ & Constraint\\
\midrule
$(1,0)$ & 1 & $s=1,\ u=0$\\
$(2,0)$ & 2 & $u\geq1$\\
$(2,1)$ & 1 & $s+u\geq1$\\
$(3,0)$ & 3 & $u=3$\\
$(3,1)$ & 2 & $u\geq1+2t\geq1$\\
$(4,0)$ & 4 & $u\geq s+2\geq2$\\
\bottomrule
\end{tabular}
\end{minipage}\end{center}

For four exposed qubits, purity and the spin-flip operation give the required constraint. Let $A_w$ sum the squared expectation values of all weight-$w$ Pauli operators in a pure four-qubit state $\rho$. Complementary reduced states have equal purity, so
\begin{equation}
 A_3=12+A_1-2A_2,\qquad A_4=3-2A_1+A_2 .
\end{equation}
The spin-flipped state $\widetilde\rho=Y^{\otimes4}\rho^T Y^{\otimes4}$ is positive, so
\begin{equation}
 0\leq\Tr(\rho\widetilde\rho)
 =\frac{1-A_1+A_2-A_3+A_4}{16}
 =\frac{A_2-A_1-2}{4}.
 \label{eq:spinflip_main}
\end{equation}
This proves Eq.~\eqref{eq:four_obstruction}, consistent with shadow-enumerator bounds and the absence of a pure four-qubit state whose two-qubit reductions are all maximally mixed~\cite{Rains1999,Higuchi2000}.

For a $(2,1)$ cluster, the exposed reduction has rank at most two, implying $(1+s+u)/4\geq1/2$. For a pure two-qubit cluster, complementary one-qubit purities give equal squared Bloch lengths $t$, so $u=3-2t\geq1$.

For three exposed qubits, equality of single-qubit and complementary two-qubit purities gives $u=3$.

For $(m,r)=(3,1)$, let $s$ be the exposed one-body sum, $t$ the protected one-body contribution, and $w$ the total exposed--protected two-body contribution. Summing equal purities of complementary pairs yields $u-w=3t-s$. The four-qubit inequality $u+w\geq s+t+2$ then gives $u\geq1+2t$.

To strengthen the overlap bound, we use a real representation in which the three Pauli error vectors at each site are orthonormal. Map $z\in\mathbb C^D$ to $\widehat z=(\operatorname{Re}z,\operatorname{Im}z)\in\mathbb R^{2D}$ and a Hermitian operator $M$ to
\begin{equation}
 \widehat M=\begin{pmatrix}\operatorname{Re}M&-\operatorname{Im}M\\
 \operatorname{Im}M&\operatorname{Re}M\end{pmatrix}.
\end{equation}
Then $\widehat u^T\widehat M\widehat v=\operatorname{Re}\langle u|M|v\rangle$, and $0\leq M\leq I$ implies $0\leq\widehat M\leq I$. This is a mathematical representation of the same measurement, without an added physical ancilla.

Let $\widehat U_j$ have columns $\widehat{\sigma_a^j\psi}$ for $a=X,Y,Z$. The Pauli anticommutation relations imply $\widehat U_j^T\widehat U_j=I_3$: the imaginary within-site overlaps disappear on taking the real part. For $j\ne l$, the operators $\sigma_a^j\sigma_b^l$ are Hermitian, so their expectations are real and
$\widehat U_j^T\widehat U_l=U_j^\dagger U_l$.
Thus realification preserves exactly the cross-site overlaps that must be bounded while making every three-column frame an isometry.

Write $e_j=\sum_a e_{ja}$ and $q_{l\to j}=\sum_a\langle\psi|\sigma_a^l M_j\sigma_a^l|\psi\rangle$. Splitting the cross-site Gram block with $\widehat M_j$ gives
\begin{align}
 \|U_j^\dagger U_l\|_F
 &\leq\|\widehat U_j^T\widehat M_j\widehat U_l\|_F\nonumber\\
 &\quad+\|\widehat U_j^T(I-\widehat M_j)\widehat U_l\|_F\nonumber\\
 &\leq\sqrt{q_{l\to j}}+\sqrt{e_j}.
 \label{eq:frame_pair}
\end{align}
For the first term, factor $\widehat M_j$ into square roots and use the operator norm on the factor containing $\widehat U_j$. Reverse the norm assignments for the second term. Each operator norm is at most one. The squared Frobenius norms of the remaining factors are $q_{l\to j}$ and $e_j$.

For a subset of $m$ exposed sites, let $E_S=\sum_{j\in S}e_j$ and $Q_S=\sum_{j,l\in S,j\ne l}q_{l\to j}$. POVM normalization gives $Q_S\leq E_S$. Minkowski's inequality on the ordered pairs therefore yields
\begin{align}
 \sqrt{2A_2(S)}
 &\leq\sqrt{Q_S}+\sqrt{(m-1)E_S}\nonumber\\
 &\leq(1+\sqrt{m-1})\sqrt{E_S}.
 \label{eq:frame_sum}
\end{align}
This proves Eq.~\eqref{eq:frame_bound_main} for any global POVM and any pure input, without requiring the sites in $S$ to form a product state.

We can now charge each missing ancilla to the summed Pauli-test failure probabilities. Every positive-deficit cluster except $(1,0)$ and $(2,1)$ satisfies $u\geq d/2$. Set $K_4=C_4/3=2h_4=4+2\sqrt3$. Because the cluster contains at most four exposed probes, Eq.~\eqref{eq:frame_bound_main} gives $d\leq2u\leq K_4E_{\rm err}^{\rm block}$. Its contribution to $E_{\rm err}$ is therefore at least $d/K_4$.

Pair the exceptional clusters. Factorization gives a weight-two sum $u_1+u_2+s_1s_2\geq1$ for each pair. The pair contains at most four exposed probes and contributes at least $2/K_4$ to $E_{\rm err}$. At most one exceptional cluster remains. Omitting its deficit gives the offset of one in the first inequality of Eq.~\eqref{eq:counting_main}.

Including overlap with the undisturbed input accounts for an unpaired exceptional cluster. The corresponding inequality is
\begin{equation}
 |\bra\psi a_j\ket\psi|^2\leq2(e_{ja}+f_j),\qquad
 f_j=\bra\psi M_j\ket\psi .
 \label{eq:null_overlap_main}
\end{equation}
For an exceptional cluster, the overlap inequalities give $s\leq2E_{\rm err}^{\rm block}+6F^{\rm block}$ and $u\leq2E_{\rm err}^{\rm block}$, where the superscript denotes sums over sites in that cluster. Because $s+u\geq1$, its deficit is bounded by $4E_{\rm err}^{\rm block}+6F^{\rm block}\leq K_4E_{\rm err}^{\rm block}+6F^{\rm block}$. Summing over all clusters gives $K_4E_{\rm err}+6F\geq N-R$. Clusters with nonpositive deficit cannot weaken either bound.

The extension to mixed conditional inputs uses linearity. Keep the POVM fixed, average the inequalities over a pure-state decomposition, and then average over the retained labels $c$. Apply the false-alarm constraint to the final average. Combining the two inequalities with $E_{\rm err}\geq0$ gives $E_{\rm err}\geq\pos{N-R-\min(1,6\alpha)}/K_4$.

To obtain the minimax bound, evaluate the protocol on the isotropic test channel
\begin{equation}
 \E_\star=(1-p_{\min})\id+
 \frac{p_{\min}}3\sum_{a=X,Y,Z}\operatorname{Ad}(\sigma_a),
 \label{eq:isotropic_main}
\end{equation}
where $\operatorname{Ad}(U)$ denotes conjugation by $U$. Substitute the error bound into its site-averaged success probability, Eq.~\eqref{eq:average_main}, to prove Theorem~\ref{thm:protection_main}. The identity/Pauli mixture is a feature of this chosen test channel. The allowed channel family still includes arbitrary coherent and nonunital disturbances.

\subsection{Definition and proof of the exposed-participation bound}
\label{app:participation}

We count the exposed probes that must belong to factors of at least five qubits in an exact product decomposition. Let $\mathsf E$ be the set of exposed probes. For a pure state $\ket\psi$, let $\Pi(\psi)$ be the set of partitions across which it factors exactly, and define
\begin{equation}
 h_5(\psi)=\min_{\pi\in\Pi(\psi)}
 \sum_{\substack{B\in\pi\\|B|\geq5}}|B\cap\mathsf E| .
 \label{eq:h5_main}
\end{equation}
The factor size $|B|$ includes exposed and protected qubits, while $|B\cap\mathsf E|$ counts only exposed probes. Minimization selects the smallest participation compatible with the state's exact factorization, so grouping independent factors cannot inflate the count. For the actual labeled input ensemble, define
\begin{equation}
 \mathcal P_5=\sum_c\lambda_c
 \inf_{\rho_c=\sum_\nu q_{\nu|c}\ket{\psi_{\nu|c}}\bra{\psi_{\nu|c}}}
 \sum_\nu q_{\nu|c}h_5(\psi_{\nu|c}),
 \label{eq:C5_main}
\end{equation}
where the infimum runs over normalized pure-state decompositions. The two labels have different roles. The index $\nu$ belongs to a mathematical decomposition; $c$ is an actual preparation record available to the detector. Accordingly, the conditional POVM stays fixed across every component in a decomposition of $\rho_c$.

For any product partition of a pure input, let $H$ count the exposed probes in factors of at least five qubits. Set those factors aside and apply the small-cluster argument to the remainder. At most $R$ ancillas remain, so
\begin{equation}
 K_4E_{\rm err}+H\geq N-R-1,\qquad
 K_4E_{\rm err}+6F+H\geq N-R .
\end{equation}
The inequalities hold for every product partition, including one with $H=h_5(\psi)$. At fixed conditional readout, $E_{\rm err}$ and $F$ are linear in the input state, so their averages are independent of the pure-state decomposition. The inequalities therefore survive the infimum in Eq.~\eqref{eq:C5_main}. Average over retained labels, apply $F\leq\alpha$, and use the minimax shortfall bound to obtain Theorem~\ref{thm:participation_main}. Throughout the argument, the detector uses its actual conditional readout and receives no mathematical decomposition label.

\subsection{Proof of the exact depth-two frontier}
\label{app:depth_two}

A single fault changes only one product factor. This lets us bound a global readout by separate decisions on the factors. It suffices to upper-bound the site-averaged response to the allowed isotropic test channel of strength $p=p_{\min}$. First consider a pure input $|\psi\rangle=\bigotimes_b|\psi_b\rangle$ and let $Q=I-|\psi\rangle\langle\psi|$. At zero false alarms every reporting effect annihilates $|\psi\rangle$. An error at a site in factor $b$, after projection with $Q$, lies in the subspace orthogonal to $|\psi_b\rangle$ in that factor, with every other factor unchanged. These subspaces are orthogonal for different $b$. Compression of a global POVM therefore defines a valid sub-POVM on each factor; there is no improvement from off-diagonal matrix elements between the subspaces.

For an all-exposed pair, local basis changes put the input in the form
\begin{equation}
 |\psi_b\rangle=\sqrt\lambda|00\rangle+\sqrt{1-\lambda}|11\rangle,
 \qquad \tfrac12\leq\lambda\leq1.
\end{equation}
Isotropic noise is invariant under those basis changes. Define $Q_b=I-|\psi_b\rangle\langle\psi_b|$ and the subnormalized nonidentity test operators
\begin{equation}
 A_j=\frac13 Q_b\sum_{a=X,Y,Z}\sigma_a^{(j)}|\psi_b\rangle\langle\psi_b|\sigma_a^{(j)}Q_b.
\end{equation}
They are diagonal in the basis $|01\rangle,|10\rangle,|\phi\rangle$, where $|\phi\rangle=\sqrt{1-\lambda}|00\rangle-\sqrt\lambda|11\rangle$. Their respective diagonal entries are
\begin{align}
 A_1 &: \tfrac23(1-\lambda),\ \tfrac23\lambda,\ \tfrac43\lambda(1-\lambda),\nonumber\\
 A_2 &: \tfrac23\lambda,\ \tfrac23(1-\lambda),\ \tfrac43\lambda(1-\lambda).
\end{align}
Maximizing a two-site decision in each sector gives a sum of site scores, divided by $p$, of at most
\begin{equation}
 \frac43\lambda+\frac43\lambda(1-\lambda)
 =\frac43(2\lambda-\lambda^2)\leq\frac43.
 \label{eq:pair_certificate}
\end{equation}
The positive diagonal operator obtained by taking the larger entry in each sector is a dual certificate for this bound.

A single pure exposed qubit contributes at most $2/3$. A factor containing one exposed and one protected qubit contributes at most one, since its three nonidentity branches have total probability one. Factors containing only protected qubits contribute nothing.

If $b$ exposed qubits are paired with protected qubits, the total normalized score is at most $b+2(N-b)/3$. Since $b\leq\min(R,N)$, dividing by $N$ proves the upper bound.

For the mixed-state converse, a mixed input of depth two has a pure product-factor decomposition. Positivity and $F=0$ force every reporting effect to annihilate every positive-weight pure component. Giving the decoder the component label can only increase its power, and the same upper bound applies to each component. Retained preparation randomization is handled in the same way.

Uniform random assignment of $\min(R,N)$ Bell ancillas and independent probes elsewhere attains the bound for every fault location and every allowed channel.

\subsection{Reduction to unrestricted cluster optima}
\label{app:cluster_reduction}

We first optimize the sum of correct-site probabilities for a fixed block input. For a normalized pure input $|\psi\rangle$ with $m$ exposed and $r$ protected qubits, write $\Pi_\psi=I-|\psi\rangle\langle\psi|$ and
\begin{equation}
 A_j(\psi)=\frac13\Pi_\psi\sum_{a=X,Y,Z}
 \sigma_a^{(j)}|\psi\rangle\langle\psi|\sigma_a^{(j)}\Pi_\psi.
 \label{eq:cluster_test}
\end{equation}
The fixed-input readout problem is the semidefinite program
\begin{equation}
 \begin{aligned}
 w(\psi)&=\max_{\substack{M_j\geq0\\\sum_jM_j\leq\Pi_\psi}}
 \sum_{j=1}^m\Tr[M_jA_j(\psi)],\\
 W^\star(m,r)&=\max_\psi w(\psi).
 \end{aligned}
 \label{eq:cluster_sdp}
\end{equation}
These are summed site scores, not site averages. Equivalently, on the orthogonal complement of $|\psi\rangle$, the fixed-state dual minimizes $\Tr Y$ subject to $Y\geq A_j(\psi)$ for every $j$. Strict feasibility on this subspace gives strong duality. The outer optimization over the input state is not convex.

For a pure product input, the argument of Appendix~\ref{app:depth_two} places each factor's non-null error contribution in a subspace orthogonal to all other factors' contributions. Compressing any global reporting POVM into these subspaces bounds the sum of site scores by $\sum_bW^\star(m_b,r_b)$. For a mixed $k$-producible input, positivity and zero false alarms force the reporting effects to annihilate every pure component of positive weight. Giving the decoder the decomposition label can only enlarge the measurement class. The bound therefore holds component by component, and then for retained classical preparation labels. Taking the isotropic channel of strength $p_{\min}$ and comparing worst-site success with the site average proves the upper bound in Proposition~\ref{prop:cluster_reduction}.

Achievability uses two randomizations: one removes the unknown channel direction, and the other equalizes the site scores. Choose optimal pure states and readouts for an optimizing list of block types. On each exposed qubit, conjugate the unknown channel by a private uniformly random single-qubit Clifford. Averaging this conjugation yields the isotropic channel with unchanged $p(\E)$. All block reports annihilate the undisturbed input, so their responses scale exactly with $p(\E)$. Uniform private permutation of the exposed labels then gives each physical site the same average block score. The controls and permutation are independent of the faulty site and channel. The resulting worst-site score equals the right-hand side of Eq.~\eqref{eq:cluster_reduction}; its minimum over the allowed channel family occurs at $p_{\min}$. This construction uses unrestricted block preparation, as permitted in the stated model, and does not assert that physically exchanging distant probes is free.

The discrete maximization can be evaluated recursively: set $\mathcal W_k(0,R)=0$ and maximize $W^\star(m,r)+\mathcal W_k(N-m,R-r)$ over $m\leq N$, $r\leq R$, and $m+r\leq k$. Protected-only factors are unnecessary and unused ancillas are allowed. The known block values are $W^\star(1,0)=2/3$, $W^\star(1,r)=1$ for $r\geq1$, $W^\star(2,0)=4/3$, and $W^\star(2,2)=2$. Thus the unresolved input optimizations at depth at most four are confined to four block types. Fixed-state semidefinite-program (SDP) bounds and searches over pure inputs can provide numerical evidence, but cannot by themselves supply an upper bound uniform over all input states.

\subsection{The sharp coefficient and its four-qubit reduction}
\label{app:sharp_constant}

The sharp coefficient is determined by the largest ancilla deficit per unit block-score loss. For each positive-deficit block define $d_{m,r}=m-r$ and $\ell_{m,r}=m-W^\star(m,r)$. The bounds in Eq.~\eqref{eq:small_block_brackets}, together with the known block values, give $\ell_{m,r}>0$ whenever $m+r\leq4$ and $m>r$. We first claim
\begin{equation}
 C_{\rm opt}=\max_{\substack{m+r\leq4\\m>r}}
 \frac{d_{m,r}}{\ell_{m,r}}.
 \label{eq:sharp_all_blocks}
\end{equation}
Let $C$ denote the right-hand side. Every positive-deficit block obeys $d_{m,r}\leq C\ell_{m,r}$. The same inequality holds trivially for a nonpositive-deficit block, since its loss is nonnegative. In any allocation using at most $R$ ancillas,
\begin{equation}
 N-R\leq\sum_b(m_b-r_b)\leq C\sum_b\ell_{m_b,r_b}.
\end{equation}
Proposition~\ref{prop:cluster_reduction} therefore proves the universal converse with coefficient $C$, including mixed inputs, retained labels and arbitrary global readout.

No smaller coefficient can suffice: a single optimizing block, with $N=m$ and $R=r$, attains normalized loss $\ell_{m,r}/m$. A smaller coefficient would exclude that allowed protocol. This proves Eq.~\eqref{eq:sharp_all_blocks}; tiling the block gives the same obstruction at larger sizes.

The ratios for $(1,0)$ and $(2,0)$ are exactly three. The unrestricted upper bounds for $(3,0)$ and $(2,1)$ imply
\begin{equation}
 \frac{3}{3-W^\star(3,0)}\leq\frac92+3\sqrt2,
 \qquad
 \frac{1}{2-W^\star(2,1)}\leq9.
\end{equation}
Both are smaller than the all-exposed tetrahedron ratio $8+4\sqrt3$. Only $(4,0)$ and $(3,1)$ can maximize Eq.~\eqref{eq:sharp_all_blocks}, proving Proposition~\ref{prop:sharp_constant}.

Furthermore, $W^\star(3,1)\leq2+\sqrt3/2$ would make its ratio at most $8+4\sqrt3$, which the $(4,0)$ ratio already reaches. This implication requires only the stated upper bound, not an exact solution of the $(3,1)$ problem.

\subsection{Graph-block certificates and allocation}
\label{app:graph_frontier}

For graph inputs, the isotropic-test discrimination problem is classical in the graph basis. A local Clifford twirl maps $\E$ to $(1-p)\id+(p/3)\sum_a\operatorname{Ad}(\sigma_a)$ with $p=p(\E)$. For a graph with adjacency matrix $\Gamma$, the syndromes of $Z_j,X_j,Y_j$ are respectively $e_j,\Gamma e_j,e_j\mathbin\oplus\Gamma e_j$. In the graph basis, the non-null isotropic-test operator for site $j$ is diagonal with entries $n_{j,s}/3$. The diagonal operator with entries $\max_j n_{j,s}/3$ for $s\ne0$, and zero at the null sector, dominates every null-compressed test operator. Its trace is the upper bound in Eq.~\eqref{eq:graph_frontier}, before division by $N$. Reporting a maximizing site in each sector attains it. Uniform private permutation of exposed labels turns the site average into a worst-site score; the response is proportional to $p$ and hence minimized at $p_{\min}$.

Let $W(e,r)$ be the largest unnormalized sum of site scores divided by $p$ over all graph inputs with $e$ exposed and $r$ protected qubits, where $1\leq e$ and $e+r\leq4$. Products with depth bound $k$ obey the exact finite recursion
\begin{equation}
 A_k(N,R)=\max_{\substack{1\leq e\leq N,\ 0\leq r\leq R\\e+r\leq k}}
 \{W(e,r)+A_k(N-e,R-r)\},
 \label{eq:graph_recursion}
\end{equation}
with $A_k(0,R)=0$. Unused protected qubits are permitted. Orthogonality of different factors' non-null error spaces makes the score additive. The optimum within this finite graph-block class is $p_{\min}A_k(N,R)/N$. The enumeration and rational certificates are described in Supplemental Material, Sec.~\ref{sm:graphs}. This optimum is restricted to the finite graph-block class.

In particular, $P_3$ and $P_4$ have respectively seven and ten distinct nonzero single-site syndromes, each with maximum site multiplicity one, yielding Eq.~\eqref{eq:small_path_scores}.

\subsection{Tetrahedron input, square-root readout and dual certificate}
\label{app:tetra}

The tetrahedron state has maximally mixed one-qubit marginals, but its two-qubit marginals are maximally mixed only within the symmetric subspace. The state in Eq.~\eqref{eq:tetra_state} has $\langle\sigma_a^{(j)}\rangle=0$ and $\langle\sigma_a^{(j)}\sigma_b^{(l)}\rangle=\delta_{ab}/3$ for $j\ne l$. These marginals illustrate the distinction between isotropic first and second collective-spin moments (order-two spin anticoherence) and maximally mixed two-qubit reductions (two-uniformity)~\cite{Baguette2014}. Write $|v_{j,a}\rangle=\sigma_a^{(j)}|T_4\rangle$ for the $m$ exposed sites. The error vectors are orthogonal to the input, and their overlaps are
\begin{equation}
 \langle v_{j,a}|v_{l,b}\rangle=\delta_{ab}(G_m)_{jl},
 \qquad G_m=\frac23 I_m+\frac13 J_m,
 \label{eq:tetra_gram}
\end{equation}
where $J_m$ is the all-ones matrix. The three Pauli-axis sectors are mutually orthogonal. Within each sector let $V_a$ have columns $v_{j,a}$ and set $W_a=V_aG_m^{-1/2}$. The columns $w_{j,a}$ of $W_a$ are orthonormal. Reporting site $j$ with the projector $M_j=\sum_a|w_{j,a}\rangle\langle w_{j,a}|$, and giving no report on the complement, defines a physical measurement.

The matrix $G_m$ has eigenvalues $(m+2)/3$ and $2/3$, with multiplicities one and $m-1$. Hence every diagonal entry of $G_m^{1/2}$ is $d_m=[\sqrt{m+2}+(m-1)\sqrt2]/(m\sqrt3)$, with $d_m^2=c_m$. Expanding a channel's Kraus operators as $K_\ell=a_\ell I+\sum_a b_{\ell a}\sigma_a$, the probability of report $l$ for a fault at $j$ is
\begin{equation}
 \Pr(l\mid\E_j)=|(G_m^{1/2})_{lj}|^2\sum_{\ell,a}|b_{\ell a}|^2.
\end{equation}
Orthogonality of the axis sectors removes all cross terms. Each column of $G_m^{1/2}$ has squared norm one, proving Eq.~\eqref{eq:tetra_response} for every qubit CPTP map.

For the converse on this input, use the isotropic test. In the orthonormal $W_a$ basis, the test operator for site $j$ in one axis sector is $G_m^{1/2}|j\rangle\langle j|G_m^{1/2}/3$. The operator $Y_a=d_mG_m^{1/2}/3$ dominates all of them: the rank-one domination condition is $\langle j|G_m^{1/2}|j\rangle/d_m=1$. Summing $Y_a$ over the three orthogonal sectors gives a dual operator with trace $m d_m^2$. Its site-averaged bound matches the construction. This proves fixed-input optimality, while leaving optimization over other input states open.

\subsection{An explicit two-exposed-qubit shared-ancilla improvement}
\label{app:shared_ancilla}

A single protected qubit can improve the discrimination of two exposed sites. Order the qubits as exposed 1, exposed 2 and protected 3. For $0\leq c\leq1$, define
\begin{equation}
 |\psi_c\rangle=\frac{\sqrt{1+c}}{2}(|000\rangle+|101\rangle)
 +\frac{\sqrt{1-c}}{2}(|010\rangle-|111\rangle).
 \label{eq:shared_ancilla_state}
\end{equation}
For this state, the two-site isotropic decision has value $w(c)=\tfrac12[\Tr(A_1+A_2)+\|A_1-A_2\|_1]$, with the operators of Eq.~\eqref{eq:cluster_test}. This expression follows by assigning the positive and negative eigenspaces of $A_1-A_2$ to sites 1 and 2 on the null-orthogonal subspace.

The decision separates into unambiguous sectors and one overlapping sector. The $X_1$ and $Y_1$ error vectors lie in the odd-parity subspace of qubits 1 and 3 and contribute $2/3$ without ambiguity. The component $(Z_2-cI)|\psi_c\rangle$ is orthogonal to that subspace and to the remaining decision sector, contributing $(1-c^2)/3$. In the orthonormal basis $(|000\rangle-|101\rangle)/\sqrt2$, $(|010\rangle+|111\rangle)/\sqrt2$, the remaining difference operator is
\begin{equation}
 A_1-A_2=\frac16
 \begin{pmatrix}-1+3c&\sqrt{1-c^2}\\\sqrt{1-c^2}&-1-3c\end{pmatrix}.
\end{equation}
Its eigenvalues are $[-1\pm\sqrt{1+8c^2}]/6$, while the trace of the corresponding $A_1+A_2$ block is one. Adding the three contributions gives
\begin{equation}
 w(c)=\frac32-\frac{c^2}{3}+\frac{\sqrt{1+8c^2}}6.
 \label{eq:shared_ancilla_curve}
\end{equation}
This family is maximized at $c^2=3/8$, where $w=41/24$. Clifford twirling followed by private random assignment of the two exposed roles gives each site success probability $41p(\E)/48$, with zero false alarms. The construction is an analytical lower bound on the unrestricted block optimum; the binary decision is optimal for the specified input, not a global certificate over all three-qubit states.

\subsection{Unrestricted small-block bounds and allocation}
\label{app:small_block_status}

The real-frame inequality gives upper bounds over all block inputs and readouts. For a pure block tested at unit isotropic strength, write $E=3(m-w)$ for the sum of Pauli-test failures. At zero false alarms, $M_j|\psi\rangle=0$, so $s\leq E$; Eq.~\eqref{eq:frame_bound_main} gives $u\leq h_m E$. Combining these inequalities with Table~\ref{tab:small_clusters} yields
\begin{align}
 \frac73\leq W^\star(3,0)&\leq4\sqrt2-3,\nonumber\\
 2+\sqrt3\leq W^\star(4,0)&\leq4-\frac{2}{3(2+\sqrt3)},\nonumber\\
 \frac{41}{24}\leq W^\star(2,1)&\leq\frac{17}{9},\nonumber\\
 \frac{13+4\sqrt{10}}9\leq W^\star(3,1)&\leq3-\frac{2}{3(3+2\sqrt2)}.
 \label{eq:small_block_brackets}
\end{align}
For $(2,1)$, use $1\leq s+u\leq(1+h_2)E=3E$. For the other three types, use respectively $u=3$, $u\geq2$ and $u\geq1$. The lower bounds are the explicit constructions. These are unrestricted analytical brackets, approximately $[2.33333,2.65685]$, $[3.73205,3.82137]$, $[1.70833,1.88889]$ and $[2.84990,2.88562]$. A gap remains in each unrestricted input optimization.

Improving the known three-exposed-qubit construction requires at least one biased one-qubit marginal in a pure input. Every pure three-qubit state with all one-qubit reductions maximally mixed is locally unitarily equivalent to the Greenberger--Horne--Zeilinger (GHZ) state. To see this, use the canonical form~\cite{Acin2000}
\begin{equation}
 \begin{aligned}
 |\psi\rangle={}&\lambda_0|000\rangle+\lambda_1e^{i\phi}|100\rangle\\
 &+\lambda_2|101\rangle+\lambda_3|110\rangle+\lambda_4|111\rangle,
 \end{aligned}
\end{equation}
with nonnegative $\lambda_i$. The first maximally mixed marginal forces $\lambda_0^2=1/2$ and $\lambda_1=0$; the other two force $\lambda_2=\lambda_3=0$, leaving $\lambda_4^2=1/2$. Local unitaries preserve the isotropic discrimination optimum. GHZ has six distinct $X/Y$ error states and one shared $Z$ error state, giving $w=7/3$. Thus any pure input improving the known $(3,0)$ construction must have at least one biased one-qubit marginal.

Supplemental Material, Sec.~\ref{sm:search}, records the numerical searches that guided these constructions.

Once the construction library is fixed, only an integer allocation problem remains. Give block $b$ deficit $d_b=m_b-r_b$ and loss $\ell_b=m_b-w_b$. The six useful positive-deficit types are $(1,0)$, $(2,0)$, $(3,0)$, $(4,0)$, $(2,1)$ and $(3,1)$, with losses
\begin{equation}
 \frac13,\quad\frac23,\quad\frac23,\quad
 2-\sqrt3,\quad\frac7{24},\quad\frac{14-4\sqrt{10}}9.
\end{equation}
Bell pairs fill the remaining exposed sites at zero loss; library blocks with nonpositive deficit cannot improve on that filling. The recurrence $L_0=0$ and $L_d=\min_{b:d_b\leq d}(L_{d-d_b}+\ell_b)$ computes the least library loss at deficit $d$, provided the reconstructed blocks use at most $N$ exposed qubits. The implementation verifies this condition throughout the plotted domain. It compares algebraic values in $\mathbb Q(\sqrt3,\sqrt{10})$ using disjoint rational enclosures, and fails if a comparison cannot be resolved. All library ratios obey $\ell_b/d_b\geq(2-\sqrt3)/4$, which also excludes larger deficits beyond the finite search domain.

At $N=1000$ and $\delta=10^{-3}$, the allocation losses satisfy
\begin{gather*}
 L_{14}=0.953946394996\ldots<1,\\
 L_{15}=1.245613061663\ldots>1.
\end{gather*}
No larger deficit meets the target. Hence 986 ancillas are sufficient and minimal within this library. At $\delta=10^{-2}$, 37 all-exposed tetrahedra and 852 Bell pairs are the best library allocation. These finite allocation certificates inherit the analytic block guarantees but do not turn the unknown $W^\star(m,r)$ into known optima.

\section{Noisy interrogation and finite-round guarantees}
\label{app:noisy_guarantees}

\subsection{Completely positive comparison under depolarizing noise}
\label{app:cp_order}

In the Pauli representation, the completely positive comparison reduces to comparing four probabilities. The background channel $\mathcal D_\eta$ has probabilities
\begin{equation}
 b_0=1-3\eta/4,\qquad b_a=\eta/4\quad(a=X,Y,Z).
\end{equation}
Composing it with $\E_\star=(1-p)\id+(p/3)\sum_a\operatorname{Ad}(\sigma_a)$, where $p=p_{\min}$, gives
\begin{equation}
 t_0=b_0-(1-\eta)p,\qquad
 t_a=b_a+(1-\eta)p/3.
 \label{eq:noise_weights}
\end{equation}
These weights are nonnegative for $0\leq p\leq1$ and obey $t_\mu\leq c_\eta b_\mu$. Thus $c_\eta\mathcal D_\eta-\mathcal D_\eta\circ\E_\star$ is completely positive.

Tensoring with the other background channels and identity channels on any ancillas yields $\mathcal C_j\leq_{\rm CP}c_\eta\mathcal C_0$ for a complete array interrogation. In particular, every input produces outputs satisfying $\rho_j\leq c_\eta\rho_0$. Each site effect therefore obeys
$\operatorname{Tr}(M_j\rho_j)\leq c_\eta\operatorname{Tr}(M_j\rho_0)$.
Sum over sites and divide by $N$. The worst-site success probability cannot exceed this average, giving $v_\eta\leq c_\eta\alpha/N$. To obtain the separate upper bound $B_\alpha$, absorb the known post-fault background into the final POVM of an ideal single interrogation. This change preserves both the achieved score and the null false-alarm probability.

The same comparison iterates through adaptive interrogations with hypothesis-independent controls. For $T$ rounds, allow arbitrary completely positive, trace-preserving controls between complete array interrogations, with feedback outcomes stored in a classical register. Suppose the states before a round satisfy $\rho_j\leq c_\eta^{t-1}\rho_0$. Positivity of the control and of $\mathcal C_j$ gives
$\mathcal C_j(\rho_j)\leq c_\eta^{t-1}\mathcal C_j(\rho_0)$, which completely positive order further bounds by $c_\eta^t\mathcal C_0(\rho_0)$. The induction remains valid with any memory register. Applying the final global POVM proves Eq.~\eqref{eq:repeated_bound}.

The argument assumes a fixed causal sequence of complete array queries, a persistent fault, memoryless background and a single overall false-alarm constraint. The budget $\alpha$ applies to the final decision, rather than separately to each round. At $\eta=1$, $c_\eta=1$ and the exposed outputs are fully depolarized, so repetition cannot restore location information.

\subsection{Achievable repeated-interrogation bounds}
\label{app:repeated}

Independent probes attain logarithmic interrogation scaling uniformly over the full fault family. In each round, choose each probe independently from the six Pauli eigenstates and measure in its preparation basis. The mean flip probability is $2p(\E)/3$ without background, and $\eta/2+(1-\eta)2p(\E)/3$ after depolarization. Thus the null mean is $q_0=\eta/2$, while a faulty site's mean exceeds it by at least
\begin{equation}
 g_{\rm ind}=\frac23(1-\eta)p_{\min}.
 \label{eq:ind_gap}
\end{equation}
Compare empirical frequencies with $q_0+g_{\rm ind}/2$ and report only a unique crossing. Fresh preparations and memoryless channels give independent trials across rounds. Hoeffding's inequality bounds each erroneous crossing or missed crossing by $e^{-Tg_{\rm ind}^2/2}$. A union bound over the null sites, or over the faulty site and its $N-1$ competitors, gives
\begin{equation}
 F\leq N e^{-Tg_{\rm ind}^2/2},\qquad
 v_{\eta,T}\geq1-N e^{-Tg_{\rm ind}^2/2}.
 \label{eq:repeated_ach}
\end{equation}
Consequently, a target probability $\beta_*$ of failing to identify the affected site is achieved with
\begin{equation}
 T\geq\frac{2}{g_{\rm ind}^2}
 \log\frac{N}{\min(\alpha,\beta_*)}.
 \label{eq:round_upper}
\end{equation}
Fresh Bell pairs obey the same bound with $q_0=3\eta/4$ and $g_{\rm Bell}=(1-\eta)p_{\min}$, at a cost of $N$ protected ancillas per round. Both protocols use $NT$ exposed-channel queries and retain only classical records between rounds. These bounds prove the achievable order, without optimizing its constant. Section~\ref{sec:finite_noise} uses the exact finite distributions for sharper budgets.

The logarithm in the denominator of Eq.~\eqref{eq:round_lower} cannot in general be linearized. Replacing $1/\log c_\eta$ by $3\eta/[4(1-\eta)p_{\min}]$ requires $4(1-\eta)p_{\min}/(3\eta)\ll1$, which excludes the small-background limit at fixed fault strength.

\subsection{Background ordering and general channel domination}
\label{app:ordering}

For post-fault local background, the adjoint-channel construction above absorbs the background into the final POVM. For pre-fault background, instead use $\rho'_c=(\mathcal D_\eta^{\otimes N}\otimes\id_A)(\rho_c)$ as the input to an ideal-background protocol. Local channels preserve mixed-state $k$-producibility: apply their Kraus operators to each pure product factor and retain the resulting mixture. The ancilla count and preparation labels are unchanged. Thus both orderings obey the ideal resource upper bounds. The same identity-Pauli branch in Eq.~\eqref{eq:background_mixture} gives the same code lower bound in either ordering.

For the repeated-interrogation converse, the chosen isotropic test channel commutes with $\mathcal D_\eta$, so the same completely positive comparison applies in either order. This reasoning concerns the test channel used in the upper bound; it does not interchange the order of an arbitrary unknown channel and background.

For a general background, Eq.~\eqref{eq:general_cp_order} is equivalent to $\mathcal B\circ\E_*\leq_{\rm CP}c\mathcal B$ by the Choi criterion. The preceding induction then proves $\rho_{j,T}\leq c^T\rho_{0,T}$. Summing the final site effects gives the stated bound. For the opposite order one must check $J(\E_*\circ\mathcal B)\leq cJ(\mathcal B)$ separately. If $J(\mathcal B)>0$, the least admissible constant is the largest eigenvalue of $J(\mathcal B)^{-1/2}J(\mathcal B\circ\E_*)J(\mathcal B)^{-1/2}$; for a singular background Choi operator, the Choi operator being compared must be supported on the background support.

\subsection{Finite-alphabet noise propagation and threshold bounds}
\label{app:finite_noise}

The circuit calculation initializes each computational-basis qubit with an independent bit error, applies the stated encoding, exposure and inverse decoding, and flips each measured bit independently. After every two-qubit gate a nonidentity two-qubit Pauli is applied uniformly among the fifteen possibilities with the specified total probability. Clifford propagation maps each such event to an additive final-record bit string. A finite convolution of these independent distributions defines $P_0$; we evaluate it in floating-point arithmetic. Exposed depolarization contributes identity with probability $1-3\eta/4$ and each Pauli with probability $\eta/4$. Optional protected-memory depolarization is propagated in the same way.

Fresh local twirling immediately before and after exposure makes the fault isotropic even when the prepared state contains the specified Pauli preparation errors. The added fault Pauli then shifts the final record, proving Eq.~\eqref{eq:syndrome_convolution}. Without circuit errors this is equivalent to enumerating all $4^5=1024$ exposed-background Pauli strings; both calculations are checked. The calculation uses finite enumeration rather than sampling or rare-event extrapolation. The comparison assumes ideal single-qubit controls and independent Pauli circuit noise; coherent gate errors, leakage and correlated drift are not included.

For a fixed site and channel, each round's site indicator is Bernoulli, with independent draws across rounds. Its count therefore has a binomial distribution. A null report requires at least one crossing. Under a fault, the event that the true site crosses and no other site crosses guarantees a correct-site report. Bounding its complement proves Eq.~\eqref{eq:finite_round_certificate}. Affinity of every one-round probability in $p$ lets endpoint evaluation bound all $p\in[p_{\min},p_{\max}]$; binomial tails are monotone in their success probability. Correlations among indicators in a round do not invalidate either union bound.

Applying the standard binomial Chernoff inequality gives Eq.~\eqref{eq:kl_rounds} and its within-block extension. Searching integer $h$ and increasing $T$ yields the first round budget certified by these inequalities. It need not be the minimum budget for the decoder or for the optimal experiment.

\subsection{Optimal selective decoding of Bernoulli records}
\label{app:optimal_decoding}

We prove Proposition~\ref{prop:optimal_count} for $0<q_0<q_1<1$. Let $n=T$ be the number of binary samples at each site; for two independent sensors per site, take $n=2T$. Under the test alternative in which site $j$ has probability $q_1$, the likelihood ratio of the entire array record is
\begin{equation}
 L_j=\left(\frac{q_1}{q_0}\right)^{K_j}
 \left(\frac{1-q_1}{1-q_0}\right)^{n-K_j}.
\end{equation}
For a decision rule with reporting probabilities $\delta_j$, the site-average success is $\mathbb E_0[\sum_j\delta_jL_j]/N$ and the null-report cost is $\mathbb E_0[\sum_j\delta_j]$. A Lagrange multiplier for the latter selects a maximizing $L_j$ whenever $\max_jL_j$ exceeds a common threshold, and selects no report below it. Boundary randomization attains any prescribed false-alarm budget. Since $L_j$ increases with $K_j$, this is the largest-count rule.

Uniform tie splitting preserves permutation symmetry, making its site-average success equal to its worst-site success. No rule can exceed its site-average optimum, which proves the worst-site upper bound.

For fixed other-site counts, the probability of correctly reporting $j$ is nondecreasing in $K_j$: it changes from zero to a possible tie probability, then to one, subject to the same monotone rejection threshold. Increasing the faulty site's Bernoulli probability therefore cannot reduce success. The rule attains its worst value at $q_1$, so the test-alternative upper bound is also the optimum for all allowed larger probabilities. For two local sensors, the same coupling works separately for each sensor's probability.

The isotropic channel at $p_{\min}$ supplies an allowed test for the full-family protocols. Under that test, the retained Clifford labels and the three nonzero Bell labels carry no information beyond the flag count. Ignoring those labels is sufficient for the uniform achievability argument.

The optimum can be evaluated without enumerating all array records. Write $b_a(k)=\Pr[\operatorname{Bin}(n,q_a)=k]$ and $G_0(k)=\sum_{l=0}^k b_0(l)$, with $G_0(-1)=0$. Given a true-site count $k$, the probability of winning after uniform tie splitting is
\begin{equation}
 H(k)=\frac{G_0(k)^N-G_0(k-1)^N}{N b_0(k)}.
 \label{eq:tie_win}
\end{equation}
To derive it, sum over the number $l$ of other sites tied at $k$, assigning the true site weight $1/(l+1)$:
\begin{equation}
 H(k)=\sum_{l=0}^{N-1}\binom{N-1}{l}
 \frac{b_0(k)^lG_0(k-1)^{N-1-l}}{l+1}.
\end{equation}
The binomial integral identity gives Eq.~\eqref{eq:tie_win}. If the maximum count is accepted above $h$, with probability $\rho$ at $h$, the exact probabilities are
\begin{align}
 F&=1-G_0(h)^N+\rho[G_0(h)^N-G_0(h-1)^N],\nonumber\\
 v&=\sum_{k=h+1}^{n}b_1(k)H(k)+\rho b_1(h)H(h).
 \label{eq:optimal_count_probability}
\end{align}
Choose $h$ and $\rho\in[0,1]$ so that $F=\alpha$. The optimal success is nondecreasing with $T$, since a decoder can ignore extra samples. Evaluating the successful round count and its predecessor therefore identifies the minimum round budget for this fixed-readout classical problem. The released calculation uses 80-digit arithmetic, repeats each accepted budget at 110 digits, and checks the formula against exact rational enumeration for small arrays. These numerical evaluations implement the analytical classical optimum; they do not establish an optimum over quantum probing strategies.

\subsection{Full-record likelihood scores and a uniform fault-strength guarantee}
\label{app:likelihood_envelope}

To obtain a guarantee uniform in fault strength, we first order the true-site score distributions. Let $Q_j(s)=\frac13\sum_aP_0(s\mathbin\oplus u_{j,a})$, so that $P_{j,p}=(1-p)P_0+pQ_j$. For the backgrounds used here $P_0(s)>0$. Set $r_j(s)=Q_j(s)/P_0(s)$. The matched score is $\ell_j(s)=\log[1-p_{\min}+p_{\min}r_j(s)]$, an increasing function of $r_j$. For any nondecreasing quantization $g_j$ of this score, the event $g_j(s)\geq x$ is an upper level set of $r_j$. Its probability is nondecreasing in $p$, because
\begin{equation}
 \frac{d}{dp}\Pr_{j,p}[g_j\geq x]
 =\mathbb E_0[(r_j-1)\mathbf1_{\{g_j\geq x\}}]\geq0.
 \label{eq:likelihood_monotonicity}
\end{equation}
The last inequality is the nonnegative covariance of two increasing functions of $r_j$, using $\mathbb E_0 r_j=1$. Independent-round coupling extends this stochastic order to the summed score. Threshold acceptance with boundary randomization is also nondecreasing, so its true-site probability is smallest at $p_{\min}$.

For a competing site $l\ne j$, monotonicity in the true fault strength need not hold. We instead dominate each tail separately. Define its one-round tail envelope
\begin{equation}
 \overline H_{l|j}(x)=\max_{p\in\{p_{\min},p_{\max}\}}
 \Pr_{j,p}[g_l\geq x].
 \label{eq:score_envelope}
\end{equation}
Every tail probability is affine in $p$, hence bounded by this maximum on the whole interval. The maximum of two tail functions is itself a tail function. Its consecutive differences define a score distribution that stochastically dominates each allowed one-round score. Taking independent copies dominates the $T$-round sum. This argument does not assume that either endpoint is least favorable for the entire wrong-site decision.

For the five-cycle we use the integer score
\begin{equation}
 g_j(s)=\left\lfloor\frac{\ell_j(s)-\min_u\ell_j(u)}{\delta}\right\rfloor,
 \qquad\delta=1/32.
\end{equation}
Its exact integer values are released with the data. Let $A_0$, $A_1$ and $\overline A_{l|j}$ denote the threshold acceptance probabilities of the summed null, weakest true-site and envelope scores. Acceptance means exceeding $h$, or equaling $h$ and accepting with probability $\rho$ using private randomness. Cyclic symmetry makes $A_0$ and $A_1$ site independent. Reporting only a unique accepted site gives
\begin{align}
 F&\leq N A_0,\nonumber\\
 v&\geq A_1-(N-5)A_0-\sum_{l\ne j}\overline A_{l|j}.
 \label{eq:likelihood_certificate}
\end{align}
No independence between scores in one block is required for these union bounds. The right-hand side holds for every $p$ in the prescribed strength interval.

The one-round probabilities are rational for the stated Pauli noise parameters. Sparse convolution evaluates the score sums, and 70-digit recomputation verifies the selected thresholds from those rational inputs. The five-cycle row of Table~\ref{tab:rare_alarm} is a sufficient guarantee for this explicit quantized decoder. It neither optimizes all full-record decisions nor saturates the relative-entropy lower bound.

\section{Physical scope and computational methods}
\label{app:physical_scope}

\subsection{Effective module channels and a known-mechanism example}
\label{app:physical}

To apply the single-probe model to a physical disturbance, the reduced evolution must define an input-independent channel. Suppose the environment of module $j$ begins in a fixed state $\tau_{B_j}$, uncorrelated with the sensor input and independent of private preparation randomness. If a CPTP map $\Lambda_{jB_j}$ acts only within that module, tracing out the environment gives
\begin{equation}
 \E_j(\rho_j)=\Tr_{B_j}\Lambda_{jB_j}(\rho_j\otimes\tau_{B_j}) .
 \label{eq:module_main}
\end{equation}
Complete positivity makes the local description valid even when the probe is entangled with other sensors. Applying the ideal resource laws also requires identity evolution on the other exposed probes and $p(\E_j)$ in the prescribed strength interval. Initial sensor--environment correlations and preparation-dependent effective channels require a different reduction.

The environment $B_j$ includes the module's data qubits and other degrees of freedom; the encoded sensors are a separate register. A CE may be a spatially broad increase in local error rates, even when faults are independent conditional on that increase. A distance-$d$ code guarantees correction of arbitrary errors up to weight $\lfloor(d-1)/2\rfloor$, but exceeding that weight does not by itself prove a decoding failure. Burst resilience depends on the noise strength, duration and geometry, together with the decoder. Surface-code simulations find regimes where increasing distance still suppresses errors when a uniform burst lasts one syndrome-extraction round and the decoder uses the corresponding noise weights~\cite{Tan2024}. We therefore use ``catastrophic'' relative to a module's operating regime, rather than as a claim that every spatially broad event defeats every code distance.

Several responses to module-level faults are complementary. Distributed outer codes can treat a diagnosed chip failure as an erasure~\cite{Xu2022}. Classical environmental sensors can flag disturbances~\cite{Orrell2021}, while syndrome-based architectures can detect bursts and adapt decoding or code allocation~\cite{Suzuki2022}. These methods exploit different observations and fault assumptions. Our reduction addresses the quantum resources needed to identify a module whose local channel is otherwise unknown. Comparing it with these alternatives requires a processor-specific model.

Knowing the mechanism can change which probe is best. For ideal amplitude damping in a known basis,
\begin{align}
 K_0&=\ket0\bra0+\sqrt{1-\gamma}\ket1\bra1,\qquad
 K_1=\sqrt\gamma\ket0\bra1,\nonumber\\
 p_{\rm AD}&=1-\frac{(1+\sqrt{1-\gamma})^2}{4}.
 \label{eq:AD_main}
\end{align}
At zero false alarms, independent probes prepared in $\ket1$ and measured in the energy basis identify a decay with probability $\gamma$. A two-uniform sensor gives correct-site probability $p_{\rm AD}$. At $\gamma=0.4$, the respective probabilities are $0.4$ and approximately $0.212702$. Knowledge of the mechanism and energy basis lets independent probes outperform the full-family construction in this narrower fault model, consistently with the minimax theorem.

The background model matters as well. The relation $\gamma=1-e^{-t/T_1}$ describes ideal Markovian relaxation during interrogation. Interpreting a fault as a change from an already noisy baseline requires specifying that baseline and revisiting the comparison.

\subsection{Fault timing during preparation and decoding}
\label{app:timing}

The sensing model places the unknown channel between encoding and decoding. This temporal assumption matters even when every gate is ideal: a fault during decoding can be assigned to a different site. Prepare the five-cycle by applying $H$ to sites $1,\ldots,5$, followed by controlled-$Z$ gates on $(1,2),(2,3),(3,4),(4,5),(5,1)$. Decode in reverse order. An $X_3$ fault inserted after the first three decoding controlled-$Z$ gates leaves only the $(2,3)$ and $(1,2)$ gates before the final Hadamards. Its propagated error is then
\begin{equation}
 X_3\longmapsto X_3Z_2\longmapsto Z_3X_2.
 \label{eq:timing_alias}
\end{equation}

The measured record is $01000$. The decoder designed for faults during exposure interprets it as $Z_2$ and reports site 2, although site 3 was disturbed.

The complete boundary audit and the distinction between sequential and parallel gate schedules are given in Supplemental Material, Sec.~\ref{sm:timing_audit}. These insertion counts have no assigned onset distribution. Applying the exposure-window resource theorem to a continuous-time detector requires a model of event onset and faults during finite-duration gates.

\subsection{Computational methods}
\label{app:computation}

The resource bounds and fixed-input certificates are proved analytically. Computations verify finite graph sectors with rational arithmetic, evaluate exact finite-alphabet decision probabilities, and compare Clifford propagation with independent state-vector calculations. Alternating state/readout searches supply feasible constructions, not unrestricted upper bounds. Supplemental Material, Secs.~\ref{sm:graphs}, \ref{sm:search} and \ref{sm:timing_audit}, describes the enumeration, search stopping rules, precision checks and circuit audit. The released code and data retain the corresponding records. These finite-case checks supplement the proofs; they neither validate a device model nor provide independent scientific review.

\clearpage
\onecolumngrid
\begin{center}
{\large\bfseries Supplemental Material for\\Quantum Codes for Distributed Quantum Fault Sensing}
\end{center}
\twocolumngrid
\makeatletter
\@removefromreset{equation}{section}
\makeatother
\setcounter{section}{0}
\setcounter{subsection}{0}
\setcounter{equation}{0}
\setcounter{figure}{0}
\setcounter{table}{0}
\renewcommand{\thesection}{S\arabic{section}}
\renewcommand{\theequation}{S\arabic{equation}}
\renewcommand{\thefigure}{S\arabic{figure}}
\renewcommand{\thetable}{S\arabic{table}}
\renewcommand{\theHsection}{supp.\arabic{section}}
\renewcommand{\theHequation}{supp.\arabic{equation}}
\renewcommand{\theHfigure}{supp.\arabic{figure}}
\renewcommand{\theHtable}{supp.\arabic{table}}

This supplement gives statistical certification, additional noise comparisons, computational records and fault audits. Appendix references and equation or figure numbers without an S prefix refer to the main paper. The analytical proofs of the main resource bounds and finite-round guarantees remain in its appendices.

\section{Resource witness and finite-sample certification}
\label{app:witness}

The resource law also certifies how many exposed probes participate in large entangled factors. We need only a specified ensemble of fault tests, rather than an estimate of the worst-case response over unknown channels. In each test, inject a uniformly selected single-site Pauli $P_{j,a}$, keep its site and axis hidden from the detector, and score whether the detector reports $j$. The mean score is
\begin{equation}
 S=\frac{1}{3N}\sum_{j,a,c}\lambda_c
 \operatorname{Tr}(M_{j|c}P_{j,a}\rho_cP_{j,a}).
 \label{eq:pauli_score}
\end{equation}
The diagnostic Pauli injections may lie outside the prescribed fault-strength interval. They test the input and readout of the sensing protocol, and null trials use that same detector. Since $E_{\rm err}=3N(1-S)$, the proof of Theorem~\ref{thm:participation_main} gives
\begin{equation}
 \mathcal P_5\geq
 \bigl[N-R-\min(1,6F)-C_4N(1-S)\bigr]_+.
 \label{eq:witness}
\end{equation}
The witness bounds the preparation resource of the tested protocol. Its interpretation requires trusted fault injection and a declared ancilla count; it does not provide device-independent certification or determine the full minimax correct-site probability.

To account for finite data, suppose that independent trials use a stationary protocol. Let $\widehat S$ and $\widehat F$ be the observed frequencies from $n_s$ Pauli trials and $n_0$ null trials. The detector must use the same input and readout across trial types: neither the trial type nor the injected label may be disclosed in a way that changes the protocol. For failure probability $\beta$, define
\begin{align}
 S_L&=\left[\widehat S-
 \sqrt{\frac{\log(2/\beta)}{2n_s}}\right]_+,\\
 F_U&=\min\!\left\{1,\widehat F+
 \sqrt{\frac{\log(2/\beta)}{2n_0}}\right\}.
 \label{eq:confidence}
\end{align}
Two one-sided Hoeffding inequalities~\cite{Hoeffding1963}, followed by a union bound, give a simultaneous lower confidence bound with coverage at least $1-\beta$ when $S_L,F_U$ are substituted in Eq.~\eqref{eq:witness}.

Calibration uncertainty adds a separate correction. If the actual and intended injection channels differ by at most $\zeta_s$ in half the diamond norm, replace $S_L$ by $[S_L-\zeta_s]_+$; an analogous null-channel bound $\zeta_0$ replaces $F_U$ by $\min(1,F_U+\zeta_0)$.

The converse constant limits the witness's sensitivity. At $R=F=0$, even a positive witness evaluated from the exact trial probabilities requires $S>1-1/C_4\simeq0.955342$. Certifying $\mathcal P_5/N\geq0.9$ requires $S\geq1-0.1/C_4\simeq0.995534$, before statistical and calibration penalties. A sharper converse would therefore improve the certificate directly. We do not claim that the constant $C_4$ is tight.

\paragraph*{Proof of the confidence bound.}

The error-counting inequalities give the population witness after substituting $E_{\rm err}=3N(1-S)$.
For independent Bernoulli trials, Hoeffding's inequality gives
\begin{align}
 \Pr(S<\widehat S-t_s)&\leq e^{-2n_st_s^2},\\
 \Pr(F>\widehat F+t_0)&\leq e^{-2n_0t_0^2}.
\end{align}
Set each right side to $\beta/2$. A union bound then ensures that both one-sided bounds hold with probability at least $1-\beta$. Because the witness increases with $S$ and decreases with $F$, substitution of Eq.~\eqref{eq:confidence}, with the stated clipping, gives a valid lower confidence bound. A trusted half-diamond-norm discrepancy $\zeta$ changes any outcome probability by at most $\zeta$, even in the presence of ancillary input systems. This establishes the calibration correction. Independent trials and stationarity justify Hoeffding's inequality; hidden labels ensure that the trials test the same protocol. The resource interpretation also requires the declared ancilla count. Drift and unrecorded side information require further assumptions or a different statistical analysis.

\section{A finite-resource separation at small background noise}
\label{app:small_noise}

A conservative bound establishes a finite-resource separation at sufficiently small nonzero background noise. Define
\begin{equation}
 \epsilon_\eta=1-(1-3\eta/4)^N.
 \label{eq:noise_penalty}
\end{equation}
This is the probability of at least one nonidentity background Pauli in the array; it is separate from the physical fault-occurrence probability. Prepare a five-cycle tiling, with $5\mid N$ and $R=0$, and report only when exactly one block produces a recognized site syndrome. Then
\begin{equation}
 F_\eta\leq\epsilon_\eta,\qquad
 v_\eta^{\rm code}\geq p_{\min}(1-\epsilon_\eta).
 \label{eq:noisy_code_lower}
\end{equation}
For comparison, absorbing the known post-fault background into the final POVM preserves input resources and all outcome probabilities. The noisy optimum therefore cannot exceed its ideal-background counterpart, so
\begin{equation}
 V^{(\eta)}_{k,\alpha}\leq B_\alpha-
 \frac{p_{\min}}{C_4N}[N-R-\kappa_\alpha]_+,
 \quad 2\leq k\leq4.
 \label{eq:noisy_resource_upper}
\end{equation}
The depth-five construction outperforms every such small-cluster protocol whenever
\begin{align}
 \epsilon_\eta&\leq\alpha,\nonumber\\
 \epsilon_\eta&<\frac{[N-R-\kappa_\alpha]_+}{C_4N}
       -\frac{(1-p_{\min})\alpha}{Np_{\min}}.
 \label{eq:noisy_separation}
\end{align}
This comparison fixes the exposed array, one query per exposed probe, the background, the full fault family and the false-alarm budget; the upper bound permits arbitrary global readout.

For $N=1000$, $p_{\min}=0.2$, $\alpha=0.005$, $R=0$ and $\eta=(4/3)[1-0.995^{1/1000}]\simeq6.68337\times10^{-6}$, the code guarantees $v_\eta\geq0.199$, whereas depth at most four gives $v_\eta\leq0.191073$. The sufficient per-probe tolerance decreases as $1/N$. This calculation does not establish an advantage at fixed nonzero $\eta$ as the array grows.

\paragraph*{Proof.}
The background has the exact convex decomposition
\begin{equation}
 \mathcal D_\eta^{\otimes N}
 =w_0\id+(1-w_0)\mathcal B,\qquad
 w_0=(1-3\eta/4)^N,
 \label{eq:background_mixture}
\end{equation}
where $\mathcal B$ is a CPTP mixture of nonidentity Pauli strings. On the identity branch the code has zero false alarms and correct-site probability $p(\E)$. Bounding the other branch's event probabilities between zero and one proves Eq.~\eqref{eq:noisy_code_lower} for arbitrary local CPTP faults. For the upper bound, move the known background to the measurement. A noisy reporting effect $M_j$ becomes the ideal-background effect $(\mathcal D_\eta^{\otimes N}\otimes\id_A)^\dagger(M_j)$. The adjoint is completely positive and unital, so it preserves the POVM constraints. This proves Eq.~\eqref{eq:noisy_resource_upper}; comparing the two bounds gives Eq.~\eqref{eq:noisy_separation}. Appendix~\ref{app:ordering} extends the argument to pre-fault background.

\begin{widetext}
\noindent\begin{minipage}{\textwidth}
\centering
\includegraphics[width=\textwidth]{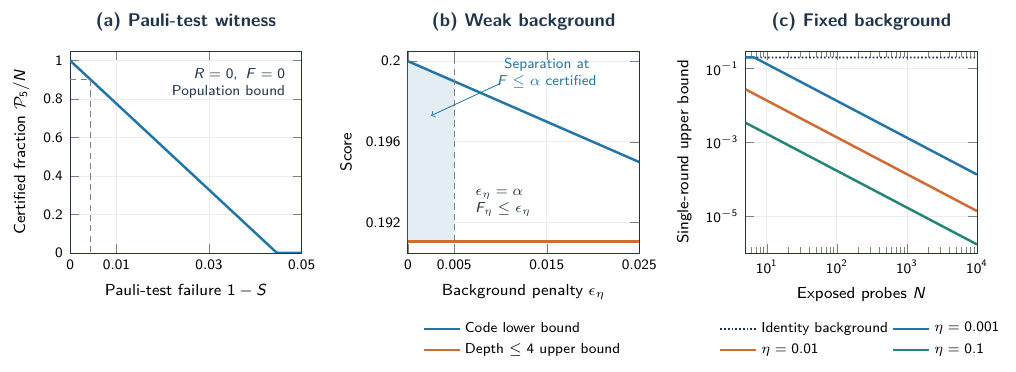}
\captionof{figure}{\textbf{A resource certificate and the cost of background noise.}
(a) Lower bound, evaluated from exact trial probabilities, on $\mathcal P_5/N$ from randomized Pauli tests at $R=F=0$. The horizontal coordinate is the Pauli-test failure probability $1-S$, which differs from the shortfall in correct-site probability. Finite-sample and calibration corrections are given in Eqs.~\eqref{eq:witness}--\eqref{eq:confidence}.
(b) At $N=1000$, $p_{\min}=0.2$, $\alpha=0.005$ and $R=0$, the five-cycle lower bound exceeds the universal upper bound for depth at most four in the shaded region. Shading also enforces $F_\eta\leq\epsilon_\eta\leq\alpha$, with boundary $\eta\simeq6.68337\times10^{-6}$. The horizontal axis is the aggregate penalty $\epsilon_\eta$, not per-probe strength $\eta$. This sufficient weak-noise window shrinks as $1/N$ in $\eta$.
(c) Universal single-interrogation upper bound $\min\{B_\alpha,c_\eta\alpha/N\}$, for any number of protected ancillas, at $p_{\min}=0.2$ and $\alpha=0.005$. The dotted curve gives the identity-background benchmark. Curves are analytical upper bounds, not optimized noisy scores or measurement data.}
\label{fig:operational}
\end{minipage}
\end{widetext}

\section{Graph enumeration and finite certificates}
\label{sm:graphs}

For each exposed count $e\geq1$ and protected count $r\geq0$ with $e+r\leq4$, the calculation enumerates all $2^{(e+r)(e+r-1)/2}$ labeled simple graphs. For each graph, it constructs the three single-Pauli syndromes at every exposed site and evaluates the sector certificate proved in Appendix~\ref{app:graph_frontier}. All scores and certificates are rational. The released data contain every enumerated graph score, a maximizing graph and its sector certificate for each block type, and the resulting $N=2,3,4$ frontiers. The allocation recursion is Eq.~\eqref{eq:graph_recursion} of the main paper. These finite certificates optimize the graph-block class; they do not optimize over arbitrary nonstabilizer inputs.

\par\medskip
\noindent\begin{minipage}{\columnwidth}
\centering
\includegraphics[width=\columnwidth]{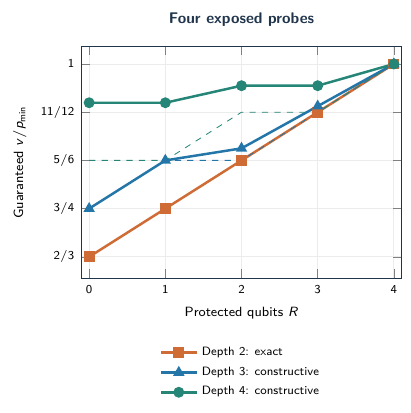}
\captionof{figure}{\textbf{Constructive scores for four exposed probes.} Zero-false-alarm scores divided by $p_{\min}$, including the nonstabilizer blocks of Table~\ref{tab:small_frontier} in the main paper. Depth two is exact over arbitrary inputs. Depth-three and depth-four curves are constructive lower bounds; their unrestricted optimality is open. Thin dashed curves show the earlier graph-block optima. The comparison uses identity background and a single interrogation.}
\label{fig:small_frontiers}
\end{minipage}
\par\medskip

\section{Numerical searches and verification records}
\label{sm:search}

Numerical searches guided the constructions, without supplying unrestricted upper bounds. The initial records cover all four unresolved block types and retain their short iteration and time limits. A targeted follow-up used the analytic tetrahedron or GHZ-equivalent path as an initial state, random perturbations, local filters that bias exposed marginals, and Haar inputs. It recorded 39 starts for $(4,0)$, 29 for $(3,1)$ and 20 for $(3,0)$, with at most 100 alternating state/readout iterations per start.

The two four-qubit runs reached their 480-second limits. Of their completed starts, 38 and 28 stopped after three successive score changes below $10^{-8}$; this is a numerical stopping rule, not a stationarity or global-optimality certificate. Seventeen three-qubit starts reached the iteration cap.

The best feasible numerical scores were $3.7320508074$ for four exposed qubits, $2.8499011822$ for three exposed qubits sharing one protected qubit, and $2.3333333331$ for three exposed qubits. All are consistent with the analytic constructions. The targeted search thus recovers the $(3,1)$ tetrahedron value missed by the initial short run, and no higher feasible score was found in these runs. Both sets of raw records retain initializations, seeds, iteration histories, stopping reasons and solver diagnostics. Repaired numerical dual bounds apply only to a fixed input and can remain loose. The analytical brackets in Eq.~\eqref{eq:small_block_brackets}, rather than numerical convergence, supply the unrestricted upper bounds.

\paragraph*{Verification methods.}

Graph frontiers use exhaustive labeled-graph enumeration and rational sector certificates. The nonstabilizer constructions have explicit analytic readouts and fixed-input dual bounds. Exact arithmetic checks the tetrahedron Gram matrix; state-vector tests check the channel responses and shared-ancilla family. Alternating optimization over a pure input and a semidefinite program for the readout guided the constructions. Released records retain seeds, solver statuses and fixed-input gaps. These searches provide feasible inputs; the analytical bounds establish the limits over all input states.

Finite Pauli propagation determines the circuit distributions. Binomial and multinomial sums give the threshold guarantees, retaining all within-block competing flags. Independent calculations compare Clifford propagation with state-vector circuits, background distributions with exhaustive Pauli strings, and selected probability sums with rational enumeration. The timing audit covers all 315 single-Pauli circuit-boundary cases for the declared sequential circuit.

Selective largest-count probabilities are checked against independent rational enumeration. The complete-syndrome guarantees are recomputed at 70-digit precision from exact one-round probabilities, and the Bernoulli budgets are checked at 80 and 110 digits. Exact rational enclosures certify the construction allocation. Classical relative entropies use the complete specified record distributions and are checked against their coarse-grained versions.

The real-frame bound is proved analytically and checked on finite complex-state and nonprojective-POVM examples. These finite-case computations supplement the analytical proofs; they neither certify unrestricted optimality nor validate a device model or provide independent scientific review.

\section{Joint threshold probabilities and a uniform channel envelope}
\label{app:joint}

The full records let us evaluate exactly the probability that only one site reaches threshold under the null hypothesis and bound the fault term uniformly, retaining correlations among sites within each block. For one null block, let $a_0$ be the probability that no site reaches threshold, and let $b_0$ be the probability that exactly one site does. For a faulty block, let $C_j$ be the probability that only its true site $j$ reaches threshold. With $L=N/m$ independent blocks,
\begin{equation}
 F=L b_0 a_0^{L-1},\qquad
 v\geq\min_j C_j a_0^{L-1}.
 \label{eq:joint_threshold}
\end{equation}
Within each block the counts are multinomial, not independent binomials. We evaluate the null probability exactly within the finite noise model. For the fault term, a pessimistic categorical distribution bounds the whole prescribed channel-strength interval: its true-site probability is the smallest allowed value and each competing-site probability is the largest. The coupling argument below proves the uniform guarantee. This is a lower bound on success, not an exact minimization over channel strength.

A round produces at most one recognized site flag in a block. If its probabilities are $q_1,\ldots,q_m$, assign the residual probability $q_\varnothing=1-\sum_iq_i$ to no recognized flag. The counts in $T$ independent rounds are multinomial. Their no-crossing probability is
\begin{equation}
 A(\bm q;T,h)=T![z^T]e^{q_\varnothing z}
       \prod_{i=1}^{m}\sum_{c=0}^{h-1}\frac{(q_i z)^c}{c!},
 \label{eq:multinomial_below}
\end{equation}
where $[z^T]$ extracts a coefficient. Conditioning on the count at site $j$ gives its unique-crossing probability
\begin{align}
 C_j(\bm q;T,h)
 &=\sum_{c=h}^{T}\binom Tc q_j^c(1-q_j)^{T-c}\nonumber\\
 &\quad\times A\left(\frac{\bm q_{-j}}{1-q_j};T-c,h\right).
 \label{eq:multinomial_unique}
\end{align}
The cases $q_j=0,1$ are interpreted directly. Set $a_0=A(\bm q_0;T,h)$ and $b_0=\sum_jC_j(\bm q_0;T,h)$. A null array reports exactly when one block has one crossing and all others have none. Under a fault, a correct report requires the true site's unique crossing and no crossing in any unaffected block. Independence between blocks proves Eq.~\eqref{eq:joint_threshold}. For $m=1$ this reduces to $F=N u_0(1-u_0)^{N-1}$ and $v=u_1(1-u_0)^{N-1}$, where $u_i=U_T(q_i,h)$.

For the five-cycle, fix the faulty site $j$. Let $q_j^*$ be the minimum of its one-round flag probability over the prescribed strength interval, and let $q_i^*$ be the maximum for each $i\ne j$. These extrema occur at the interval endpoints. In every reported parameter set, $\sum_iq_i^*\leq1$, so these numbers define a categorical distribution. Start with a round drawn from this distribution. Reassign excess competing flags and, if needed, no-flag outcomes to the true site or to no flag. This produces the distribution at any allowed strength while never decreasing the true-site count or increasing a competing count. Independent coupling across rounds makes the unique-true-crossing event monotone. Hence $C_j(\bm q^*;T,h)$ lower-bounds the success factor uniformly in the channel strength. If the envelope probabilities sum to more than one, this construction is unavailable and must not be used.

The implementation evaluates Eq.~\eqref{eq:multinomial_below} by nonnegative coefficient convolution, avoiding subtraction of nearly equal probabilities at rare false-alarm budgets. Exact rational enumeration of 72 small count problems checks the implementation. Checks at selected channel strengths further test the envelope; its uniform guarantee follows from the coupling proof.

\section{Complete circuit-timing audit}
\label{sm:timing_audit}

The audit uses the sequential preparation and reverse decoding circuit specified in Appendix~\ref{app:timing} of the main paper. Its explicit wrong-site example is Eq.~\eqref{eq:timing_alias}.

We enumerate the three Pauli faults at each of five sites at all 21 boundaries of this 20-gate sequence, and check all 315 cases with an independent state-vector calculation. All fifteen faults at the intended exposure boundary are located correctly.

Restricting to the ten other boundaries within the controlled-$Z$ preparation and decoding stages gives 150 cases: 96 correct reports, 16 wrong-site reports, 16 unrecognized nonzero records and 22 zero records.

These unweighted counts classify insertion points and Pauli axes; they are separate from physical event probabilities. The other boundaries test a different hypothesis family. They demonstrate why a channel inserted during exposure cannot represent arbitrary event onset. A continuous-time detector would need a timing model, gate durations, and a decoder that distinguishes or deliberately groups the resulting histories. The present audit assigns no onset distribution and establishes no detection latency.

The same cycle admits three parallel controlled-$Z$ layers if the device can run disjoint edges simultaneously: $\{(1,2),(3,4)\}$, $\{(2,3),(4,5)\}$ and $\{(5,1)\}$. The sequential boundary counts above do not describe that schedule. Converting either schedule into event probabilities requires an onset distribution, fault support and dynamics during gates; weighting insertion points by gate duration alone does not specify these ingredients. A fault during a finite-duration gate need not propagate as a single Pauli. Applying the exposure-window resource theorem to a continuous-time detector therefore requires an explicit model of event onset and faults during gates.

\section{Two-site Pauli faults in a five-cycle}
\label{sm:multiple}

Multiple affected modules create a different ambiguity. Among the 90 weight-two Paulis on a five-cycle, 30 produce recognized single-site syndromes, all reporting a third, unaffected site; the other 60 are unrecognized. Zero null false alarms does not prevent these wrong-site reports. Separating all Pauli sectors through weight two would resolve the supports of those Pauli errors. A resource law for general two-site channels also requires a specified fault family and report convention. Predicting a device's detection latency or total fault-mitigation cost further requires models of event onset, multiple affected modules and intermodule operations.

The $\binom52 3^2=90$ weight-two Pauli operators, counted without phase, have syndromes obtained by adding the two single-site syndromes modulo two. Enumeration gives the counts above. This classifies Pauli operators without assigning probabilities to a physical two-site channel: the 30 recognized records report wrong sites under the enlarged fault family, rather than false alarms under the no-fault hypothesis.


\begin{thebibliography}{44}
\bibitem{Xu2022} Q. Xu, A. Seif, H. Yan, N. Mannucci, B. O. Sane, R. Van Meter, A. N. Cleland, and L. Jiang, Distributed Quantum Error Correction for Chip-Level Catastrophic Errors, \href{https://doi.org/10.1103/PhysRevLett.129.240502}{Phys. Rev. Lett. \textbf{129}, 240502 (2022)}.

\bibitem{Suzuki2022} Y. Suzuki, T. Sugiyama, T. Arai, W. Liao, K. Inoue, and T. Tanimoto, Q3DE: A fault-tolerant quantum computer architecture for multi-bit burst errors by cosmic rays, in \emph{Proceedings of the 55th IEEE/ACM International Symposium on Microarchitecture} (2022), pp.~1110--1125, \href{https://doi.org/10.1109/MICRO56248.2022.00079}{doi:10.1109/MICRO56248.2022.00079}.

\bibitem{Vepsalainen2020} A. P. Veps{\"a}l{\"a}inen et al., Impact of ionizing radiation on superconducting qubit coherence, \href{https://doi.org/10.1038/s41586-020-2619-8}{Nature \textbf{584}, 551 (2020)}.

\bibitem{Wilen2021} C. D. Wilen et al., Correlated charge noise and relaxation errors in superconducting qubits, \href{https://doi.org/10.1038/s41586-021-03557-5}{Nature \textbf{594}, 369 (2021)}.

\bibitem{McEwen2022} M. McEwen et al., Resolving catastrophic error bursts from cosmic rays in large arrays of superconducting qubits, \href{https://doi.org/10.1038/s41567-021-01432-8}{Nat. Phys. \textbf{18}, 107 (2022)}.

\bibitem{Harrington2025} P. M. Harrington et al., Synchronous detection of cosmic rays and correlated errors in superconducting qubit arrays, \href{https://doi.org/10.1038/s41467-025-61385-x}{Nat. Commun. \textbf{16}, 6428 (2025)}.

\bibitem{Li2025} X. Li et al., Cosmic-ray-induced correlated errors in superconducting qubit array, \href{https://doi.org/10.1038/s41467-025-59778-z}{Nat. Commun. \textbf{16}, 4677 (2025)}.

\bibitem{Google2025} Google Quantum AI and Collaborators, Quantum error correction below the surface code threshold, \href{https://doi.org/10.1038/s41586-024-08449-y}{Nature \textbf{638}, 920--926 (2025)}.

\bibitem{Tan2024} S. J. S. Tan, C. A. Pattison, M. McEwen, and J. Preskill, Resilience of the surface code to error bursts, \href{https://doi.org/10.1103/vwtl-hwfx}{Phys. Rev. A \textbf{113}, 022450 (2026)}.

\bibitem{Orrell2021} J. L. Orrell and B. Loer, Sensor-Assisted Fault Mitigation in Quantum Computation, \href{https://doi.org/10.1103/PhysRevApplied.16.024025}{Phys. Rev. Applied \textbf{16}, 024025 (2021)}.

\bibitem{Kelly2016} J. Kelly et al., Scalable in situ qubit calibration during repetitive error detection, \href{https://doi.org/10.1103/PhysRevA.94.032321}{Phys. Rev. A \textbf{94}, 032321 (2016)}.

\bibitem{Wagner2021} T. Wagner, H. Kampermann, D. Bru{\ss}, and M. Kliesch, Optimal noise estimation from syndrome statistics of quantum codes, Phys. Rev. Research \textbf{3}, 013292 (2021).

\bibitem{Wagner2022} T. Wagner, H. Kampermann, D. Bru{\ss}, and M. Kliesch, Pauli channels can be estimated from syndrome measurements in quantum error correction, \href{https://doi.org/10.22331/q-2022-09-19-809}{Quantum \textbf{6}, 809 (2022)}.

\bibitem{Harper2020} R. Harper, S. T. Flammia, and J. J. Wallman, Efficient learning of quantum noise, \href{https://doi.org/10.1038/s41567-020-0992-8}{Nat. Phys. \textbf{16}, 1184 (2020)}.

\bibitem{Degen2017} C. L. Degen, F. Reinhard, and P. Cappellaro, Quantum sensing, \href{https://doi.org/10.1103/RevModPhys.89.035002}{Rev. Mod. Phys. \textbf{89}, 035002 (2017)}.

\bibitem{Proctor2018} T. J. Proctor, P. A. Knott, and J. A. Dunningham, Multiparameter Estimation in Networked Quantum Sensors, \href{https://doi.org/10.1103/PhysRevLett.120.080501}{Phys. Rev. Lett. \textbf{120}, 080501 (2018)}.

\bibitem{Eldredge2018} Z. Eldredge, M. Foss-Feig, J. A. Gross, S. L. Rolston, and A. V. Gorshkov, Optimal and secure measurement protocols for quantum sensor networks, \href{https://doi.org/10.1103/PhysRevA.97.042337}{Phys. Rev. A \textbf{97}, 042337 (2018)}.

\bibitem{Zhuang2020} Q. Zhuang and S. Pirandola, Ultimate Limits for Multiple Quantum Channel Discrimination, Phys. Rev. Lett. \textbf{125}, 080505 (2020).

\bibitem{Skotiniotis2024} M. Skotiniotis, S. Llorens, R. Hotz, J. Calsamiglia, and R. Mu\~noz-Tapia, Identification of malfunctioning quantum devices, Phys. Rev. Research \textbf{6}, 033329 (2024).

\bibitem{Pereira2021} J. L. Pereira, L. Banchi, Q. Zhuang, and S. Pirandola, Idler-free channel position finding, Phys. Rev. A \textbf{103}, 042614 (2021).

\bibitem{Harney2021} C. Harney and S. Pirandola, Idler-free multi-channel discrimination via multipartite probe states, npj Quantum Inf. \textbf{7}, 153 (2021).

\bibitem{HarneyDynamic2021} C. Harney and S. Pirandola, Analytical bounds for dynamic multi-channel discrimination, \href{https://doi.org/10.1103/PhysRevA.104.032402}{Phys. Rev. A \textbf{104}, 032402 (2021)}.

\bibitem{Piani2009} M. Piani and J. Watrous, All Entangled States Are Useful for Channel Discrimination, \href{https://doi.org/10.1103/PhysRevLett.102.250501}{Phys. Rev. Lett. \textbf{102}, 250501 (2009)}.

\bibitem{Tan2008} S.-H. Tan, B. I. Erkmen, V. Giovannetti, S. Guha, S. Lloyd, L. Maccone, S. Pirandola, and J. H. Shapiro, Quantum Illumination with Gaussian States, \href{https://doi.org/10.1103/PhysRevLett.101.253601}{Phys. Rev. Lett. \textbf{101}, 253601 (2008)}.

\bibitem{Knill1997} E. Knill and R. Laflamme, Theory of quantum error-correcting codes, \href{https://doi.org/10.1103/PhysRevA.55.900}{Phys. Rev. A \textbf{55}, 900 (1997)}.

\bibitem{Laflamme1996} R. Laflamme, C. Miquel, J. P. Paz, and W. H. Zurek, Perfect Quantum Error Correcting Code, Phys. Rev. Lett. \textbf{77}, 198 (1996).

\bibitem{Scott2004} A. J. Scott, Multipartite entanglement, quantum-error-correcting codes, and entangling power of quantum evolutions, Phys. Rev. A \textbf{69}, 052330 (2004).

\bibitem{Hyllus2012} P. Hyllus et al., Fisher information and multiparticle entanglement, \href{https://doi.org/10.1103/PhysRevA.85.022321}{Phys. Rev. A \textbf{85}, 022321 (2012)}.

\bibitem{Toth2012} G. T\'oth, Multipartite entanglement and high-precision metrology, \href{https://doi.org/10.1103/PhysRevA.85.022322}{Phys. Rev. A \textbf{85}, 022322 (2012)}.

\bibitem{Kessler2014} E. M. Kessler, I. Lovchinsky, A. O. Sushkov, and M. D. Lukin, Quantum Error Correction for Metrology, Phys. Rev. Lett. \textbf{112}, 150802 (2014).

\bibitem{Zhou2018} S. Zhou, M. Zhang, J. Preskill, and L. Jiang, Achieving the Heisenberg limit in quantum metrology using quantum error correction, \href{https://doi.org/10.1038/s41467-017-02510-3}{Nat. Commun. \textbf{9}, 78 (2018)}.

\bibitem{Layden2019} D. Layden, S. Zhou, P. Cappellaro, and L. Jiang, Ancilla-Free Quantum Error Correction Codes for Quantum Metrology, \href{https://doi.org/10.1103/PhysRevLett.122.040502}{Phys. Rev. Lett. \textbf{122}, 040502 (2019)}.

\bibitem{Brun2006} T. Brun, I. Devetak, and M.-H. Hsieh, Correcting Quantum Errors with Entanglement, \href{https://doi.org/10.1126/science.1131563}{Science \textbf{314}, 436 (2006)}.

\bibitem{Wilde2020} M. M. Wilde, M. Berta, C. Hirche, and E. Kaur, Amortized channel divergence for asymptotic quantum channel discrimination, \href{https://doi.org/10.1007/s11005-020-01297-7}{Lett. Math. Phys. \textbf{110}, 2277 (2020)}.

\bibitem{Nielsen2002} M. A. Nielsen, A simple formula for the average gate fidelity of a quantum dynamical operation, Phys. Lett. A \textbf{303}, 249 (2002).

\bibitem{Hein2006} M. Hein, W. D\"ur, J. Eisert, R. Raussendorf, M. Van den Nest, and H. J. Briegel, Entanglement in graph states and its applications, in \emph{Quantum Computers, Algorithms and Chaos}, Proceedings of the International School of Physics ``Enrico Fermi'', Vol.~162 (IOS Press, 2006), pp.~115--218, \href{https://doi.org/10.3254/978-1-61499-018-5-115}{doi:10.3254/978-1-61499-018-5-115}.

\bibitem{Baguette2014} D. Baguette, T. Bastin, and J. Martin, Multiqubit symmetric states with maximally mixed one-qubit reductions, \href{https://doi.org/10.1103/PhysRevA.90.032314}{Phys. Rev. A \textbf{90}, 032314 (2014)}.

\bibitem{Supplemental} See the Supplemental Material included after the references for finite-sample resource certification, weak-background comparisons, graph enumeration, numerical search and verification records, joint-indicator decoding, and circuit-timing and two-site fault audits.

\bibitem{ZhuangAUD2020} Q. Zhuang, Ultimate limits of approximate unambiguous discrimination, \href{https://doi.org/10.1103/PhysRevResearch.2.043276}{Phys. Rev. Research \textbf{2}, 043276 (2020)}.

\bibitem{Chefles1998} A. Chefles, Unambiguous discrimination between linearly independent quantum states, \href{https://doi.org/10.1016/S0375-9601(98)00064-4}{Phys. Lett. A \textbf{239}, 339 (1998)}.

\bibitem{Rains1999} E. M. Rains, Quantum shadow enumerators, IEEE Trans. Inf. Theory \textbf{45}, 2361 (1999).

\bibitem{Higuchi2000} A. Higuchi and A. Sudbery, How entangled can two couples get?, Phys. Lett. A \textbf{273}, 213 (2000).

\bibitem{Acin2000} A. Ac\'in, A. Andrianov, L. Costa, E. Jan\'e, J. I. Latorre, and R. Tarrach, Generalized Schmidt Decomposition and Classification of Three-Quantum-Bit States, \href{https://doi.org/10.1103/PhysRevLett.85.1560}{Phys. Rev. Lett. \textbf{85}, 1560 (2000)}.

\bibitem{Hoeffding1963} W. Hoeffding, Probability inequalities for sums of bounded random variables, \href{https://doi.org/10.1080/01621459.1963.10500830}{J. Am. Stat. Assoc. \textbf{58}, 13 (1963)}.
\end{thebibliography}
\end{document}